\documentclass[12pt,aps,prd,preprint,showpacs,nofootinbib,showkeys]{revtex4-1}

\pdfoutput=1

\usepackage{multirow}
\usepackage{graphicx}
\usepackage{amsmath}
\usepackage{amssymb}
\usepackage{tabularx}
\usepackage[colorlinks=false,linktocpage=true]{hyperref}
\usepackage[usenames]{color}

\newcommand{\beq}{\begin{equation}}
\newcommand{\eeq}{\end{equation}}
\newcommand{\bqa}{\begin{eqnarray}}
\newcommand{\eqa}{\end{eqnarray}}
\newcommand{\bsq}{\begin{subequations}}
\newcommand{\esq}{\end{subequations}}

\begin{document}
 

\title{Charmonia and Bottomonia in a Magnetic Field}
\author{Jeremy Alford}
\author{Michael Strickland} 
\affiliation{Department of Physics, Kent State University, Kent, OH 44242 United States}

\date{\today}

\begin{abstract}
We study the effect of a static homogeneous external magnetic field on charmonium and bottomonium states.
In an external magnetic field, quarkonium states do not have a conserved center-of-mass momentum.  Instead there is a new conserved quantity called the pseudomomentum which takes into account the Lorentz force on the particles in the system.
When written in terms of the pseudomomentum, the internal and center-of-mass motions do not decouple and, as a result, the properties of quarkonia depend on the states' center-of-mass momentum. 
We analyze the behavior of heavy particle-antiparticle pairs subject to an external magnetic field assuming a three dimensional harmonic potential and Cornell potential plus spin-spin interaction.  In the case of the Cornell potential, we also take into account the mixing of the $\eta_c$ and $J/\psi$ states and $\eta_b$ and $\Upsilon$ states due to the background magnetic field.  We then numerically calculate the dependence of the masses and mixing fractions on the magnitude of the background magnetic field and center-of-mass momentum of the state.
\end{abstract}



\maketitle 


\section{Introduction}
\label{sect:intro}

The behavior of matter subject to magnetic fields has been a subject of interest for
physicists for quite some time.  Already over one hundred years ago Pieter Zeeman showed
that an external magnetic field affected the spectrum of light emitted by a flame \cite{Zeeman:1897-1,Zeeman:1897-2,Zeeman:1897-3}.
In recent years there has been considerable attention focused on the
question of what happens to matter in the presence of extremely strong magnetic fields.
There are at least two situations in which extremely strong magnetic fields are expected to be 
generated:  (1) During early times after non-central heavy ion collisions one expects 
$B \sim m_\pi^2 \sim 10^{18}$ Gauss at energies probed by the Relativistic Heavy Ion Collider (RHIC) and 
$B \sim 15 \, m_\pi^2 \sim 1.5 \times 10^{19}$ Gauss at Large Hadron Collider (LHC) energies 
\cite{Fukushima:2008xe,Skokov:2009qp,Fukushima:2010vw,Voronyuk:2011jd,Deng:2012pc,Tuchin:2013ie} 
and (2) in the interior of magnetars, which are a class of neutron stars which possess 
magnetic fields on the order of $10^{18}$-$10^{19}$ Gauss \cite{1992ApJ...392L...9D}.  In this paper, 
we study the behavior of charmonium and bottomonium states subject 
to magnetic fields with an eye towards applications to the phenomenology of relativistic heavy ion collisions.

Interest in the effects of strong magnetic fields in heavy ion collisions has become a hot topic recently following the prediction of a non-trivial quantum chromodynamics (QCD) effect dubbed ``the chiral magnetic effect'' which stems from small P- and CP-odd interactions inducing an electromagnetic current when a quark-gluon plasma (QGP) is placed in an external magnetic field \cite{Fukushima:2008xe}.  There has been much work related to this in recent years and in addition it has been shown how to self-consistently take into account this effect through Berry curvature flux in the presence of a magnetic field \cite{PhysRevLett.109.181602,Chen:2012ca}.  The existence of such high magnetic  fields has also prompted many research groups to study how the finite temperature deconfinement and chiral phase transitions are affected 
by the presence of a strong background magnetic field.  These studies have included direct numerical investigations 
using lattice QCD \cite{DElia:2010nq,DElia:2011zu,Bali:2011qj,Bali:2012zg,Bali:2013esa} 
and theoretical investigations using a variety of methods including, for example, perturbative QCD studies,
model studies, and string-theory inspired anti-de Sitter/conformal field theory (AdS/CFT) correspondence 
studies \cite{Agasian:2008tb,Fraga:2008qn,Mizher:2010zb,Fukushima:2010fe,Gatto:2010qs,Gatto:2010pt,%
Preis:2010cq,Preis:2011sp,Andersen:2011ip,Erdmenger:2011bw,Gorbar:2011jd,Skokov:2011ib,Kashiwa:2011js,%
Fraga:2012fs,Fraga:2012ev,Fraga:2012rr,Andersen:2012dz,Shovkovy:2012zn,Preis:2012fh,%
Ferrari:2012yw,Fayazbakhsh:2012vr,Fukushima:2012xw,dePaoli:2012cz,Alexandre:2000jc}.  

In this paper, we consider the effects of magnetic fields on heavy quarkonium states, focussing on 1s charmonium and bottomonium states.  The physics of quantum mechanical bound states in a background magnetic field is complicated by the fact that in a background magnetic field the center-of-mass (COM) momentum is not a conserved quantity due to the breaking of translational invariance.  Instead one must take into account the Lorentz force on the constituents and construct a quantity called the COM pseudomomentum \cite{PhysRev.79.549,PhysRev.81.222,PhysRev.85.259,Carter1967,Gorkov1968,PhysRevA.4.59,Avron1978431,WunnerHerold1979,PavlovVerevkin1980244,Avron1981,PhysRevA.20.2287,Herold1981,RevModPhys.55.109,PhysRevA.52.1837}.  However, in practice one finds that, even after expressing the Hamiltonian in 
terms of the pseudomomentum, it is not possible to factorize the Hamiltonian into free COM motion
plus decoupled internal motion.  As a result, the spectrum of a bound state in background magnetic field depends
on the COM momentum of the system.  To the best of our knowledge, the first theoretical consideration of motional effects was by Lamb \cite{PhysRev.85.259} and as we will show this effect is related to the so-called motional Stark effect.

In this paper we investigate the effect of strong magnetic fields on heavy quarkonium states including 
such motional effects.  Heavy quarkonium is a nice test bed for QCD since heavy quark
states are dominated by short distance physics and can be treated using heavy quark effective theory \cite{Georgi:1990um}.
Based on such effective theories of QCD, non-relativistic quarkonium states can be reliably
described. Their binding energies are much smaller than the quark mass $m_q\gg\Lambda_{\rm QCD}$ ($q=c,b$), and their sizes are much larger than $1/m_q$. Since the velocity of the quarks in the bound state is small ($v\ll c$), quarkonium can be understood in terms of
non-relativistic potential models such as the Cornell potential which can
be derived directly from QCD using effective field theory 
\cite{Eichten:1979ms,Lucha:1991vn,Brambilla:2004jw}.  

We present numerical calculations using a Cornell potential supplemented by a spin-spin interaction which allows for a splitting between the spin-singlet and spin-triplet states.  This study contributes to ongoing discussions of the effect of strong magnetic fields on QCD bound states \cite{Marasinghe:2011bt,Tuchin:2011cg,Yang:2011cz,Simonov:2012if,Simonov:2013jpa,Andreichikov:2013zba,Machado:2013rta}.  Apart from the long range interactions, which are fundamentally different, the physics of heavy quarkonium is very similar to positronium \cite{Pirenne:1946,Pirenne:1947,Pirenne:1947-2,Pirenne:1947-3,Berestetskii:1949,Berestetskii:1949-2,PhysRev.84.858,PhysRev.84.601,PhysRev.85.1047,PhysRev.87.848,PhysRev.94.904,Hughes:1957zz,PhysRevA.8.625}.\footnote{For a nice review of positronium physics see Ref.~\cite{Berko1980}.}  In addition to motional effects \cite{PhysRevA.8.625}, it is also necessary to take into account the hyperfine mixing in the background magnetic field.  In positronium this results in a change of the spin-singlet and spin-triplet energy eigenvalues and ``quenching'' of 
ortho-positronium $3\gamma$ decays \cite{PhysRev.94.904,PhysRev.84.601,PhysRev.85.1047}.  Analogous effects occur in quarkonium and we present quantitative calculations of the effect including a realistic heavy quark interaction potential.  In addition, we present exact analytic formulas which can be obtained assuming a harmonic interaction between the constituents.  The harmonic interaction results are used for purposes of discussion and also to check the numerical methods which are applied in the more realistic case.

The structure of the paper is as follows.  In Sec.~\ref{sect:pseudomomentum} we introduce the 
pseudomomentum. In Sec.~\ref{sect:twoparticles} we discuss the application to two particle
states and then specialize to the case of particle-antiparticle states.  In Sec.~\ref{sect:mse}
we discuss the relation of the pseudopotential derived in the previous section to the motional
Stark effect.  In Sec.~\ref{sec:comsubtract} we discuss the prescription we use to subtract
the energy associated with the center of mass motion.  In Sec.~\ref{sec:spinmix} we discuss
the mixing of the spin singlet and triplet states in the presence of a magnetic field.  In
Sec.~\ref{sect:tune} we present the potential we use for our final results.  In Sec.~\ref{sect:results}
we present our numerical results.  In Sec.~\ref{sect:conc} we present our conclusions and outlook
for the future.  In three appendices we collect details of the inter-quark potential used and resulting 
spectra, our numerical method for solving the 3d Schr\"odinger equation, and an investigation of what 
happens to a harmonic state with a given center of mass momenta when a magnetic field is turned on suddenly.


\section{Particle in a constant magnetic field}
\label{sect:pseudomomentum}

We begin with the basics by introducing the pseudomomentum in the context of a single classical non-relativistic charged spin one-half particle in a background magnetic field.  
As we will demonstrate, unlike the particle momentum, the pseudomomentum
is conserved since it takes into account the Lorentz force on the particle.
The classical non-relativistic Hamiltonian for a particle in a constant magnetic field can be written
\beq
{\cal H} = \frac{1}{2m} [{\bf p} - q {\bf A}({\bf r})]^2 + V({\bf r})
- \boldsymbol{\mu} \cdot{\bf B} + m \, ,
\eeq
where $m$ is the rest mass of the particle and we assume ${\bf B}({\bf x}) = (0,0,B)$ which, in symmetric gauge, can be expressed in terms of the vector potential
${\bf A}({\bf r}) = \frac{1}{2} {\bf B}\times{\bf r} = \frac{1}{2} B (-y,x,0)$.

We can apply Hamilton's equations to derive the equation of motion
\bqa
- \frac{\partial {\cal H}}{\partial r_i} &=& \dot{p}_i \, , \nonumber \\
\frac{\partial {\cal H}}{\partial p_i} &=& \dot{r}_i \, .
\eqa
The second Hamilton equation gives $m \dot{r}_i = p_i - q A_i$
which allows us to solve for the canonical momentum, $p_i = m \dot{r}_i + q A_i$.
Using this, we can evaluate the full time derivative of the canonical momentum
\bqa
\dot{p}_i &=&  
m \ddot{r}_i  +  q \left( \frac{\partial A_i}{\partial t} 
+ \frac{d r_j}{d t} \frac{\partial A_i}{\partial r_j} \right)\, , \nonumber \\
&=&  
m \ddot{r}_i  + q  \dot{r}_j \frac{\partial A_i}{\partial r_j} \, ,
\eqa
where, in going from the first to second line we have used the fact that the vector potential is static in the
case under consideration.  The right hand side of the first Hamilton equation gives
\bqa
- \frac{\partial {\cal H}}{\partial r_i} &=& \frac{1}{m} \left({\bf p} - q {\bf A}({\bf r})\right)\cdot\left(q\frac{\partial {\bf A}}{\partial r_i}\right) - \frac{\partial V}{\partial r_i} \, , \nonumber \\
&=& q \dot{r}_j \frac{\partial A_j }{\partial r_i} - \frac{\partial V}{\partial r_i} \, .
\eqa
Equating the two sides we obtain
\beq
m \ddot{r}_i  = q \dot{r}_j \frac{\partial A_j }{\partial r_i} - q  \dot{r}_j \frac{\partial A_i}{\partial r_j} - \frac{\partial V}{\partial r_i}
\eeq
Using ${\bf v} \times {\bf B} =  \dot{\bf r} \times (\nabla \times {\bf A}) = \nabla(\dot{\bf r}\cdot{\bf A})-
 (\dot{\bf r}\cdot\nabla){\bf A}$ we can rewrite this as
\beq
m \ddot{\bf r}  = q \dot{\bf r} \times {\bf B} - \nabla V \, .
\eeq
In the case that the there is no potential, $V=0$, we have only the Lorentz force acting on the particle
\beq
m \ddot{\bf r}  = q \dot{\bf r} \times {\bf B} \, ,
\eeq
which shows that the momentum is not conserved in a constant magnetic field, as expected; however,
we can introduce a quantity which is conserved called the {\em pseudomomentum}, $\boldsymbol{\cal K}$,
\bqa
\boldsymbol{\cal K} &=& m \dot{\bf r} + q {\bf B} \times {\bf r} \, , \nonumber \\
&=& {\bf p} + \frac{q}{2} {\bf B} \times {\bf r} \, , \nonumber \\
&=& {\bf p} + q {\bf A} \, ,
\eqa
such that the equation of motion can be expressed as
\beq
\frac{d}{dt} \boldsymbol{\cal K} = 0 \, .
\eeq

\section{Two coupled particles in a constant magnetic field}
\label{sect:twoparticles}

We next consider the case of two particles subject to a translationally invariant potential
in non-relativistic quantum mechanics.  We will follow closely the treatment found in
Ref.~\cite{Herold1981}.
The Hamiltonian operator for two particles in a constant magnetic field can be written as
\beq
{\cal H} = \frac{1}{2m_1} [{\bf p}_1 - q_1 {\bf A}({\bf r}_1)]^2 
+ \frac{1}{2m_2} [{\bf p}_2 - q_2 {\bf A}({\bf r}_2)]^2  + V({\bf r}_1 - {\bf r}_2)
- \boldsymbol{\mu} \cdot{\bf B} + m_1 + m_2 \, ,
\eeq
where $\boldsymbol{\mu} = \boldsymbol{\mu}_1 + \boldsymbol{\mu}_2$ is the sum of the two
particles' magnetic moments and ${\bf B}({\bf x}) = (0,0,B)$, which can be expressed in terms of the vector potential
${\bf A}({\bf r}) = \frac{1}{2} {\bf B}\times{\bf r} = \frac{1}{2} B (-y,x,0)$ in symmetric gauge.  As usual,
${\bf p}_i = - i \nabla$ is the momentum operator for the $i^{\rm th}$ particle.
As in the previous section,
one finds that the COM momentum of the system is no longer conserved.  This is due
to the breaking of translational invariance by the vector potential (changing the origin changes
{\bf A}).  In order to preserve translational invariance in a constant magnetic field  an 
additional gauge transformation is required.  This can be achieved by introducing the 
generalized pseudomomentum operator \cite{Herold1981}
\beq
{\cal K}_k = \sum_{j=1}^{2} \left( - i \frac{\partial}{\partial x_{jk}} 
- q_j \int_0^{{\bf r}_j} \frac{\partial{\bf A}}{\partial x_k} \cdot d{\bf r}\right) ,
\eeq
where $k = 1,2,3$ denotes cartesian components.  Integrating and discarding a
constant one obtains
\beq
{\boldsymbol{\cal K}} = \sum_{j=1}^{2} \left( {\bf p}_j - q_j {\bf A}_j + q_j {\bf B} \times {\bf r}_j \right) ,
\eeq
In the gauge used herein we have ${\bf A}({\bf r}) = \frac{1}{2} {\bf B}\times{\bf r}$
which allows us to simplify this to
\bqa
{\boldsymbol{\cal K}} &=& \sum_{j=1}^{2} \left( {\bf p}_j  + \frac{1}{2} q_j {\bf B} \times {\bf r}_j \right) , \nonumber \\
&=& \sum_{j=1}^{2} \left( {\bf p}_j  + q_j {\bf A}_j \right) ,
\eqa
which is the generalization of the one particle case obtained in the previous section.  One can 
verify explicitly that the pseudomomentum operator commutes with the Hamiltonian
\beq
[{\boldsymbol{\cal K}},{\cal H}] = 0 \, .
\eeq
One can also compute the commutator of two components of ${\boldsymbol{\cal K}}$ in which
case one obtains
\beq
[{\cal K}_k,{\cal K}_l] = -i {\varepsilon}_{klm} B_m \left( \sum_{j=1}^2 q_j \right) \, ,
\eeq
which means that one will only be able to determine all components of ${\boldsymbol{\cal K}}$ simultaneously
for a electric charge neutral system.

\subsection{Two particles with equal and opposite charge}

In this section we specialize to the case that $q_1 = -q_2 = q$.  To proceed we introduce center of mass and relative coordinates
\bqa
{\bf R} &=& \frac{m_1  {\bf r}_1 + m_2  {\bf r}_2}{M} \, , \nonumber \\
{\bf r} &=& {\bf r}_1 - {\bf r}_2 \, ,
\eqa
where $M = m_1 + m_2$.   As is standard, we can express the individual positions as
\bqa
{\bf r}_1 &=&  {\bf R} + \frac{\mu}{m_1} {\bf r} \, , \nonumber \\
{\bf r}_2 &=&  {\bf R} - \frac{\mu}{m_2} {\bf r} \, ,
\eqa
where $\mu = m_1 m_2/M$ is the reduced mass.

This allows us to simplify the pseudomomentum operator
\bqa
{\boldsymbol{\cal K}} &=& \sum_{j=1}^{2} \left( {\bf p}_j  + \frac{1}{2} q_j  {\bf B} \times {\bf r}_j \right) ,
\nonumber \\
&=& - i \left( \frac{\partial}{\partial{\bf r}_1} + \frac{\partial}{\partial{\bf r}_2} \right) 
+ \frac{1}{2} q {\bf B} \times ( {\bf r}_1 - {\bf r}_2) \, ,
\nonumber \\
&=& - i  \frac{\partial}{\partial{\bf R}}
+ \frac{1}{2} q {\bf B} \times {\bf r} \, .
\eqa
Since the system is neutral, the full two-particle eigenfunctions $\Phi$ of the Hamiltonian are simultaneous
eigenfunctions of all components ${\cal K}_i$ of the pseudomomentum with eigenvalues $K_i$.  This 
allows us to factorize the full wavefunction
\beq
\Phi({\bf R},{\bf r}) =
\exp\left[ i \left( {\bf K} - \frac{1}{2} q {\bf B} \times {\bf r}\right) \cdot {\bf R} \right] \Psi({\bf r}) 
\equiv \phi({\bf R},{\bf r}) \Psi({\bf r}) 
\eeq
which satisfies ${\cal K}_j \Phi = K_j \Phi$ by construction.

Expanding out the two-particle Hamiltonian one finds the ``relative'' Hamiltonian
\bqa
{\cal H}_{\rm rel} &=& \frac{{\bf K}^2}{2 M} - \frac{q}{M} ({\bf K}\times{\bf B})\cdot{\bf r} 
+ \frac{{\bf p}^2}{2 \mu} + \frac{q}{2}\left(\frac{1}{m_1} 
- \frac{1}{m_2}\right) {\bf B}\cdot ({\bf r}\times{\bf p}) \nonumber \\
&& \hspace{1cm} + \frac{q^2}{8 \mu} ({\bf  B}\times{\bf r})^2 + V({\bf r}) 
- \boldsymbol{\mu} \cdot{\bf B} + m_1 + m_2 \, ,
\eqa
where ${\bf p} = - i \nabla$ is the relative momentum operator and one has the new eigenvalue
equation ${\cal H}_{\rm rel} \Psi({\bf r}) = E \Psi({\bf r})$.  Note that, unlike the case 
without the external field, the energy eigenvalue $E$ depends on the value of ${\bf K}$ through
coupling in the second term and not only through the term ${\bf K}^2/2M$.

\subsection{Heavy-light system}

In the limit that $m_2 \rightarrow \infty$ while holding $m_1$ fixed, we have $M \rightarrow \infty$
and $\mu = m_1 \equiv m$ and we obtain
\bqa
{\cal H}_{\rm rel} &=&  
\frac{{\bf p}^2}{2 m} - \frac{q}{2 m} {\bf B}\cdot ({\bf r}\times{\bf p}) 
+ \frac{q^2}{8 m} ({\bf  B}\times{\bf r})^2 + V({\bf r}) 
- \boldsymbol{\mu} \cdot{\bf B} + m \, ,
\eqa
where we have discarded the infinite constant $m_2$ in this case.
Recalling that ${\bf A} = \frac{1}{2} {\bf B} \times {\bf r} = \frac{1}{2} B (-y,x,0)$ one has
$({\bf B} \times {\bf r})^2 = B^2\rho^2$ and using ${\bf B}\cdot ({\bf r}\times{\bf p}) =
({\bf B} \times {\bf r}) \cdot {\bf p} = \rho B  p_\phi = - i B \partial_\phi$ we
obtain
\bqa
{\cal H}_{\rm rel} &=&  
- \frac{1}{2 m} \nabla^2 +  \frac{i}{2} \omega_c \frac{\partial}{\partial \phi} 
+ \frac{m \omega_c^2}{8} \rho^2 + V({\bf r}) 
- \boldsymbol{\mu} \cdot{\bf B} + m \, ,
\eqa
where $\omega_c = q B/m$.  This is the standard non-relativistic Hamiltonian for a spin-one-half particle subject to a potential $V$ and an external magnetic field.

\subsection{Particle-antiparticle pair}

For a bound state consisting of a particle-antiparticle pair we have $m_1=m_2=m$, $M=2m$, and $\mu = m/2$.
In this case the relative Hamiltonian simplifies to
\bqa
{\cal H}_{\rm rel} &=& \frac{{\bf K}^2}{2 M} - \frac{q}{M} ({\bf K}\times{\bf B})\cdot{\bf r} 
- \frac{\nabla^2}{2 \mu}  
+ \frac{q^2}{8 \mu} ({\bf  B}\times{\bf r})^2 + V({\bf r}) 
- \boldsymbol{\mu} \cdot{\bf B} + M\, .
\eqa
Next we decompose ${\bf K} = K_x \hat{\bf x} + K_y \hat{\bf y} + K_z \hat{\bf z}$ and simplify the expression above to obtain
\bqa
{\cal H}_{\rm rel} &=& \frac{{\bf K}^2}{2 M} + \frac{qB}{4\mu} K_x y - \frac{qB}{4\mu} K_y x 
- \frac{\nabla^2}{2\mu}  
+ \frac{q^2 B^2}{8\mu} \rho^2 + V({\bf r}) 
- \boldsymbol{\mu} \cdot{\bf B}  + M \, .
\label{eq:relham}
\eqa

\subsubsection{Relation between the pseudomomentum and kinetic center-of-mass momentum}

We now derive a general relation between the pseudomomentum and kinetic COM momentum.
The COM kinetic momentum of the system is given by 
\bqa
{\bf P}_{\rm kinetic} &=& \sum_{j} \left( - i \frac{\partial}{\partial {\bf r}_j} - q_j {\bf A}_j  \right) \, , \nonumber \\
&=& - i \frac{\partial}{\partial {\bf R}} - \frac{1}{2} q {\bf B} \times {\bf r} \, .
\eqa
Therefore, we have
\beq
\langle {\bf P}_{\rm kinetic} \rangle = \frac{\int_{\bf R}\int_{\bf r} \Phi^*
\left[ -i \frac{\partial}{\partial {\bf R}} - \frac{1}{2} q {\bf B} \times {\bf r} \right] \! \Phi }
{\int_{\bf R}\int_{\bf r} \Phi^*\Phi} \, .
\eeq
Using
\beq
-i \frac{\partial}{\partial {\bf R}} \Phi = \left({\bf K} - \frac{1}{2} q {\bf B} \times {\bf r}\right) \Phi \, ,
\eeq
one finds
\beq
\langle {\bf P}_{\rm kinetic} \rangle = {\bf K} - q {\bf B} \times \langle {\bf r} \rangle \, .
\label{eq:comrel}
\eeq

\subsection{Particle-antiparticle pair with a harmonic interaction}

We now specialize to the case that the potential is harmonic in which case the wave functions and energy levels can be obtained analytically.  Some of the results contained in this subsection were first obtained explicitly by Herold et al~\cite{Herold1981}.  We repeat the derivation here in order to use them as a basis for discussion of the COM momentum dependence of the energy.  We also use this case as a check for our numerics since it can be solved analytically.

Using the general relative Hamiltonian for a particle-antiparticle pair (\ref{eq:relham}) and $V({\bf x}) =  \frac{1}{2} k  {\bf x}^2 =
\frac{1}{2} \mu \omega_0^2 (x^2+y^2+z^2)$ we have
\bqa
{\cal H}_{\rm rel} &=& \frac{{\bf K}^2}{2 M} - \frac{\nabla^2}{2\mu}  
+  \frac{1}{2} \mu \left(\omega_0^2 + \frac{\omega_c^2}{4} \right) (x^2+y^2) 
- \frac{\omega_c K_y}{4}  x + \frac{\omega_c K_x}{4}  y
+ \frac{1}{2} \mu \omega_0^2 z^2
- \boldsymbol{\mu} \cdot{\bf B}  + M \, , \nonumber \\
 &=& \frac{{\bf K}^2}{2 M} - \frac{\nabla^2}{2\mu} 
+ \frac{1}{2} a (x^2+y^2) - bx + cy
+ \frac{1}{2} d z^2
- \boldsymbol{\mu} \cdot{\bf B}  + M \, ,
\eqa
where $\omega_c = q B/\mu$, $\boldsymbol{\mu} = \boldsymbol{\mu}_1 + \boldsymbol{\mu}_2$,
$a = \mu (\omega_0^2 +\omega_c^2/4)$,
$b =  \omega_c K_y/4$,
$c =  \omega_c K_x/4$, and
$d = \mu \omega_0^2$.
We can rewrite the third, fourth, and fifth terms using
\beq
\frac{1}{2} a (x^2+y^2) - b x + c y = \frac{1}{2} a \left[\left(x- \frac{b}{a}\right)^2 + \left(y+\frac{c}{a}\right)^2 \right] - \frac{1}{2a}(b^2+c^2) \, .
\eeq
We can simplify things further by making use of a constant coordinate shift $\bar{x} \equiv x- b/a$ and $\bar{y} \equiv y+ c/a$.
\bqa
{\cal H}_{\rm rel} &=& \frac{{\bf K}^2}{2M} - \frac{\nabla^2}{2\mu} 
+ \frac{1}{2} a (\bar{x}^2+\bar{y}^2)
+ \frac{1}{2} d z^2
- \frac{1}{2a}(b^2+c^2)
- \boldsymbol{\mu} \cdot{\bf B} + M
\eqa
which suggests that we use cylindrical coordinates with $\bar{x} = \rho \cos\phi$, $\bar{y} = \rho \sin\phi$, and 
$z =z $.  After this, the eigenvalue equation ${\cal H}_{\rm rel} \Psi = E \Psi$ becomes
\beq
\left( - \frac{\nabla^2}{2\mu} + \frac{1}{2} a \rho^2 + \frac{1}{2} c z^2 \right) \Psi({\bf r}) 
= \left(E - \frac{{\bf K}^2}{2M} + \frac{b^2}{2a} + \boldsymbol{\mu} \cdot{\bf B} + M \right)  \Psi({\bf r})
\, .
\eeq
Factorizing the relative wavefuction as $\Psi({\bf r}) = e^{i \ell \phi} Z(z) \psi(\rho)$ we find
\beq
\left( - \frac{\partial^2}{\partial \rho^2} - \frac{1}{\rho}\frac{\partial}{\partial\rho} + \frac{|\ell|^2}{\rho^2}
+ \alpha^4 \rho^2 \right) \psi
= 2 \mu \lambda \psi
\, ,
\eeq
where $\alpha^2 = \sqrt{\mu a} = \mu \sqrt{\omega_0^2 + \omega_c^2/4}$, $\lambda = E - E_z - {\bf K}^2/2M + (b^2+c^2)/2a + \boldsymbol{\mu} \cdot{\bf B} + M$ and $E_z$ is the eigenvalue of the separated $z$-equation
\beq
\left( - \frac{\partial^2}{\partial z^2} +  \gamma^4 z^2 \right) Z = 2 \mu E_z Z \, ,
\eeq
where $\gamma = (\mu c)^{1/4} = \sqrt{\mu \omega_0}$ which has a solution
\beq
Z = N\,e^{-\frac{1}{2} \gamma^2 z^2}\,H_{n_z}(\gamma z) ,
\eeq
and energy eigenvalue
\beq
E_z =  \left(n_z+ \frac{1}{2}\right) \omega_0 \, .
\eeq
Convergence as $\rho\rightarrow\infty$ requires
\beq
\lambda = \frac{\alpha^2}{\mu} (2n_\perp + 1 + |\ell|) = (2n_\perp + 1 + |\ell|)  \sqrt{\omega_0^2+\frac{\omega_c^2}{4}}
\, .
\eeq
Solving for $E$ we obtain the energy eigenvalues for the system
\beq
E_{{\bf K},n_\perp n_z \ell} = \frac{{\bf K}^2}{2M} - \frac{\omega_c^2 (K_x^2+K_y^2)}{32 \mu (\omega_0^2 + \omega_c^2/4)}
+  \left(n_z+ \frac{1}{2}\right) \omega_0  
+ (2n_\perp + 1 + |\ell|)  \sqrt{\omega_0^2+\frac{\omega_c^2}{4}}
- \boldsymbol{\mu}\cdot{\bf B} + M \, .
\label{eq:harmen}
\eeq
We can now write the full two-particle wave function 
\beq
\Phi_{{\bf K},{n_\perp}n_z \ell}({\bf R},{\bf r}) = {\cal N} \,   
\rho^{|\ell|} e^{i \ell \phi} e^{-\frac{1}{2} \gamma^2 z^2}\,e^{-\frac{1}{2}\alpha^2 \rho^2}\,
H_{n_z}(\gamma z)  L_{n_\perp}^{|\ell|}(\alpha^2\rho^2) 
e^{i \left( {\bf K} - \frac{1}{2} q {\bf B} \times {\bf r}\right) \cdot {\bf R}} \, ,
\label{eq:2pwfnc}
\eeq
where ${\cal N}$ is a normalization constant and
\bqa
\omega_c &=& \frac{qB}{\mu} \, , \nonumber \\
\alpha^2 &=&  \mu \sqrt{ \omega_0^2 + \frac{\omega_c^2}{4} } \, , \nonumber \\
\beta &=&  \frac{\omega_c}{4\mu(\omega_0^2 +\omega_c^2/4)} \, , \nonumber \\
\gamma^2 &=&  \mu \omega_0 \, , \nonumber \\
\rho^2 &=& \left(x - \beta K_y \right)^2 + \left(y + \beta K_x \right)^2 \, , \nonumber \\
\phi &=& \arctan\!\left(\frac{y + \beta K_x}{x - \beta K_y}\right) .
\eqa

\noindent
{\em Center-of-mass Kinetic Momentum}
\vspace{1mm}

Using this we can analytically compute the relationship between the pseudomomentum and the COM kinetic momentum of the state.
Using Eq.~(\ref{eq:comrel}) and
\beq
{\bf B}\times{\bf r} = B (-y,x,0) = B \left(-\rho\sin\phi + \frac{c}{a},\rho\cos\phi + \frac{b}{a},0 \right) \, ,
\eeq
one finds in this case
\beq
\langle {\bf P}_{\rm kinetic} \rangle = {\bf K} - \frac{q B c}{a}\hat{\bf x} - \frac{q B b}{a}\hat{\bf y} \, .
\eeq
Plugging in the definitions of $a$, $b$, and $c$ we obtain
\beq
\langle {\bf P}_{\rm kinetic} \rangle = \left(\frac{4 \omega_0^2}{4 \omega_0^2 + \omega_c^2} K_x, \frac{4 \omega_0^2}{4 \omega_0^2 + \omega_c^2} K_y, K_z \right) ,
\label{eq:pkinharm}
\eeq
As we can explicitly see from this expression, the components of the kinetic COM momentum do not directly correspond to the pseudomomentum components.  We note that in App.~\ref{app:sudden} we derive this formula in a different manner by assuming a time-dependent magnetic field which turns on rapidly. 

\section{Relation to the Motional Stark Effect}
\label{sect:mse}

One way to intuitively understand the result obtained in Eq.~(\ref{eq:relham}) is try to 
derive it in a different manner.  We can instead try to write down the non-relativistic 
Hamiltonian in the COM rest frame.  This step is self-contradictory since, as
we have pointed out previously, the COM momentum is not a conserved quantity in the
presence of an external magnetic field; however, let's ignore this for the time being
and assume that we can, in fact, boost to the rest frame of the state.  As before, we
assume that the magnetic field points in the $z$-direction and as a result the dynamics
in the $z$ direction is straightforward.  Putting the system at rest in the $z$-direction
and assuming that we can also hold it at rest in the $y$-direction we need only consider boosts in the $x$ direction with $v_x = P_x/M$.  In the lab frame there is only a magnetic field.  In the co-moving frame there will be both electric and magnetic fields.  Using the standard 
transformation laws for electric and magnetic fields one finds
\beq \label{eq:ltrans}
\begin{aligned}
E_x^\prime &= 0 \, , \nonumber \\
E_y^\prime &= - \gamma v_x B \approx - v_x B \, , \nonumber \\
E_z^\prime &= 0 \, ,
\end{aligned}
\qquad
\begin{aligned}
B_x^\prime &= 0 \, , \nonumber \\
B_y^\prime &= 0 \, , \nonumber \\
B_z^\prime &= \gamma B \approx B \, ,
\end{aligned}
\eeq
where for the terms with $\approx$ appearing we have discarded terms of the order $v_x^2$ and higher.

As we can see from the relations above, if we boost to the rest frame of the state, there 
is an additional electric interaction of the form $H_{\rm electric}^\prime = - q {\bf E}^\prime \cdot {\bf r}$ where ${\bf r} = {\bf r}_1 - {\bf r}_2$ is the relative position.  Using
the expressions above one finds trivially
\beq
H_{\rm electric}^\prime = \frac{qB}{4\mu} P_x y \, ,
\eeq
where we have used the fact that for a particle-antiparticle system $M = 4\mu$.  As we
can see this is precisely the ``extra term'' in Eq.~(\ref{eq:relham}) (assuming $P_x = K_x$ and $P_y = K_y = 0$).  If we had allowed for a general direction for the COM momentum, we would have generated both terms.  So we can see that the physical origin of these terms is, in fact,
the motional Stark effect; however, deriving things in this manner we have blurred the 
important distinction between ${\bf P}$ and ${\bf K}$, where only the latter is a conserved quantity.  In
what follows we will simply use Eq.~(\ref{eq:relham}) since it is the correct expression.

\section{Center-of-mass Kinetic Energy Subtraction}
\label{sec:comsubtract}

Since the energy of a particle-antiparticle state in the presence of a magnetic field has 
a non-trivial dependence on the pseudomomentum quantum number ${\bf K}$, one has to specify the precise
manner in which the energy associated with the COM motion is subtracted from
the total energy.  Our prescription for doing this is to subtract $\langle {\bf P}_{\rm kinetic} \rangle^2/2M$
where $M = m_1 + m_2 = 2 m_q$ from the total energy with $\langle {\bf P}_{\rm kinetic}\rangle$ computed via Eq.~(\ref{eq:comrel}).  

As a concrete example, let's return to the case of a harmonic interaction.  As demonstrated 
in the previous section this can be computed analytically in the case of a harmonic 
interaction.  Taking Eq.~(\ref{eq:harmen}) and subtracting $\langle {\bf P}_{\rm kinetic} \rangle^2/2M$
with $\langle {\bf P}_{\rm kinetic}\rangle$ given in Eq.~(\ref{eq:pkinharm}) we obtain
\bqa
\tilde{E}_{{\bf K},n_\perp n_z \ell} &=& E_{{\bf K},n_\perp n_z \ell} - \frac{\langle {\bf P}_{\rm kinetic} \rangle^2}{2M} \, ,
\nonumber \\
&=& \frac{2 \omega_c^2 \omega_0^2 (K_x^2+K_y^2)}{M(\omega_c^2+4\omega_0^2)^2} 
+  \left(n_z+ \frac{1}{2}\right) \omega_0  
+ (2n_\perp + 1 + |\ell|)  \sqrt{\omega_0^2+\frac{\omega_c^2}{4}}
- \boldsymbol{\mu}\cdot{\bf B} + M \, . 
\nonumber \\
\label{eq:harmensub}
\eqa
As we can see from this expression, as $B \rightarrow 0$ the dependence
of the COM-subtracted energy on the COM pseudomomentum 
vanishes as it should; however, for non-vanishing background
magnetic field, there is still a residual dependence on the components of the
pseudomomentum which are perpendicular to the background magnetic field.
In the case of the harmonic interaction, we are able to obtain the answer analytically.  
In cases other than the simple harmonic interaction, it may not be possible to obtain analytic expressions. 
Absent analytic expressions for the energy and necessary expectation values, one must perform the subtraction prescribed in this
section numerically.  

\section{Quarkonium spin-mixing}
\label{sec:spinmix}

Thus far we have not discussed the effects of the magnetic field-spin coupling for 
particle-antiparticle states.  In this respect states like the $J/\psi$ and $\Upsilon$ 
are similar to positronium (see e.g. \cite{Berko1980,Karshenboim:2003vs} and references 
therein).  We now review the mixing of the single and triplet states for completeness.  
The Hamiltonian can be written in the form
\beq
{\hat H} = {\hat H}_0 - {\boldsymbol \mu}\cdot{\bf B} \, ,
\eeq
where ${\hat H}_0$ collects all terms which depend on the spatial coordinates and
\bqa
{\boldsymbol \mu} &=& {\boldsymbol \mu}_q + {\boldsymbol \mu}_{\bar q}
\nonumber \\
&=& g^- \mu_q {\bf S}_q + g^+ \mu_q {\bf S}_{\bar q}
\nonumber \\
&=& \frac{1}{2} g' \mu_q ( {\boldsymbol \sigma}^- - {\boldsymbol \sigma}^+ ) \, ,
\eqa
where $\mu_q = Q/2m_q$ is the quark magneton and in going from the second to third lines we have used $g^- = - g^+ = g'$.  Herein, we ignore effects of the anomalous magnetic moment and take $g' = 2$.
The coupled spin states to be considered are 
\bqa
|11\rangle &=& |\!\uparrow\uparrow\,\rangle \, , \nonumber \\
|1{-}1\rangle &=& |\!\downarrow\downarrow\,\rangle \, , \nonumber \\
|10\rangle &=& \frac{1}{\sqrt{2}} \big( |\!\uparrow\downarrow\,\rangle + |\!\downarrow\uparrow\,\rangle \big) \, , \nonumber \\
|00\rangle &=& \frac{1}{\sqrt{2}} \big( |\!\uparrow\downarrow\,\rangle - |\!\downarrow\uparrow\,\rangle \big) \, .
\eqa
In the case of $c\bar{c}$ states, the 1s triplet and singlet states correspond to the $J/\psi$ and the $\eta_c$, respectively.  For $b\bar{b}$ states the 1s triplet and singlet states correspond to the $\Upsilon(1s)$ and $\eta_b$, respectively.
Without a spin-spin interaction, these states would be degenerate.  With a spin-spin interaction, the triplet and single states split.  In vacuum, the charmonium $1s$ splitting is approximately $\Delta E =$ 113 MeV and for bottomonium it is approximately $\Delta E =$ 63 MeV.

In the presence of a magnetic field there is mixing between some of these spin states.  One can easily verify that 
\bqa
(\sigma_z^+ - \sigma_z^-)|1{\pm}1\rangle &=& 0 \, , \nonumber \\
(\sigma_z^+ - \sigma_z^-)|10\rangle &=& 2\,|00\rangle \, , \nonumber \\
(\sigma_z^+ - \sigma_z^-)|00\rangle &=& 2\,|10\rangle \, .
\eqa
From this we see that there is no magnetic field effect on the $|1{\pm}1\rangle$ spin states but there will be mixing between the $|00\rangle$ and $|10\rangle$ spin states.  To determine the effect of the mixing we need only consider the two-dimensional eigensystem for the $|00\rangle$ and $|01\rangle$ states.  To proceed we shift the zero of the Hamiltonian energy to the midpoint between the unperturbed singlet and triplet states and write an effective Hamiltonian of the form
\bqa
H_{\rm eff} =
\frac{\Delta E}{2} \left(
\begin{array}{cc}
1 & \chi  \\
\chi & {-}1
\end{array}
\right) ,
\eqa 
where
\beq
\chi = \frac{2 g' \mu_q B}{\Delta E} \, .
\eeq
The resulting eigenstates can be expressed as
\beq
|\psi_\pm\rangle = \frac{1}{\sqrt{1+\varepsilon_{\pm}^2}} \Big( |00\rangle + \varepsilon_{\pm} |10\rangle\Big) \, , 
\label{eq:spineigenstates}
\eeq
with $\varepsilon_\pm \equiv (1 \pm \sqrt{1+\chi^2})/\chi$.  One can verify that the states are orthogonal and normalized.
The energy shifts of the states relative to the case of no spin-magnetic field effects taken into account are
\beq
\Delta E_\pm = \pm \frac{\Delta E}{2} (\sqrt{1+\chi^2}-1) \, .
\eeq
As a result, we see an increase in the energy of the $|10\rangle$ state and a decrease in the energy of the $|00\rangle$ state.  In what follows we will indicate the two degenerate unmixed triplet states with a superscript $\pm$, e.g. $J/\psi^\pm$ and $\Upsilon^\pm$, and the spin-mixed triplet state with a superscript $0$, e.g. $J/\psi^0$ and $\Upsilon^0$.  To close this section we note that in addition to the shifts in the energy levels, the state-mixing implies that e.g. some portion of $|10\rangle$ decays will be suppressed, instead appearing as decays with an invariant mass given by the energy of the $|\psi_-\rangle$ state.  This will cause suppression of e.g. $\Upsilon$ decays to lepton pairs and turn on decays of the $\eta_b$ to lepton pairs.  This would manifest itself experimentally as a reduction in dilepton yields at the $\Upsilon$ mass along and the appearance of a peak at the mass of the $\eta_b$.  The suppression described above is similar to the experimentally
well-known magnetic field suppression of the ortho-positronium $3\gamma$ decays \cite{PhysRev.94.904,PhysRev.84.601,PhysRev.85.1047}.

\section{Hamiltonian Reduction and Choice of Potential}
\label{sect:tune}

In some cases, such as the case of a harmonic interaction, the energies and wave functions
can be solved for analytically; however, in most cases this is not possible.  In these cases
it is necessary to solve the Schr\"odinger equation numerically.  In practice, we can subtract
out any terms which are independent of the position from Eq.~(\ref{eq:relham}).  In addition,
if the potential still possesses azimuthal symmetry we can set either $K_x$ or $K_y$ to zero
by rotating the coordinate system appropriately.  We choose herein to set $K_y$ to zero.  
The resulting Hamiltonian which is used in the numerical solutions is then of the form
\bqa
{\cal H}_{\rm rel}^\prime &=& - \frac{\nabla^2}{2\mu} + \frac{qB}{4\mu} K_x y 
+ \frac{q^2 B^2}{8\mu} \rho^2 + V({\bf r}) \, .
\label{eq:relhamsub}
\eqa
After numerical solution using (\ref{eq:relhamsub}) the constant terms can be added back in
manually in order to obtain the full energy eigenvalues.

For the charmonium and bottomonium states considered in this manuscript we use a Cornell potential plus
a spin-spin interaction with a separate spin-spin potential
\begin{equation}
V(r) = -\frac{4}{3} \frac{\alpha_s}{r} + \sigma r + ({\bf S}_1\cdot{\bf S}_2) \, V_s(r) \, .
\label{eq:potmodel}
\end{equation}
The expectation value $\langle {\bf S}_1\cdot{\bf S}_2 \rangle$ reduces to -3/4 for the singlet state and 1/4 for the triplet states.  For the spin potential $V_s(r)$ we use a form found from fits to the charm spin-spin
potential in lattice studies \cite{Kawanai:2011jt}
\begin{equation}
V_s(r) = \gamma e^{-\beta r} \, .
\label{eq:sspotmodel}
\end{equation}
For charmonia, the constants $\gamma$ and $\beta$ above were fit to lattice data in Ref.~\cite{Kawanai:2011jt}.  They found $\gamma = 0.825$ GeV and $\beta = 1.982$ GeV.  In this paper we allow for variation of $\gamma$.  For both charm and bottom states we will hold $\beta$ fixed to the value from Ref.~\cite{Kawanai:2011jt}, but we adjust the amplitude $\gamma$ in order to reproduce the experimentally measured splittings using Eq.~(\ref{eq:potmodel}) as the interaction potential.  We present the resulting parameter sets and the corresponding $B=0$ spectra of charmonium and bottomonium states in App.~\ref{app:tune}.  For the bottom system we present a single ``tuning'' which reproduces all states through the $\Upsilon(3s)$ with a maximum error of 0.22\%.  In the charm sector, we consider two different tunings:  (a) the bottom-tuned parameter set just described (see Tables~\ref{tab:bottomspec} and \ref{tab:charmspec} in App.~\ref{app:tune})  and (b) a charm-tuned parameter set which reproduces the masses of the $c\bar{c}$ $1s$ and $2s$ states with a maximum error of 1.3\% (see Table~\ref{tab:charmspec2} in App.~\ref{app:tune}).

We note that the interaction potential and the non-derivative terms in (\ref{eq:relhamsub}) can be combined into a
``pseudopotential'' of the form
\bqa
V_{\rm pseudo}(r) &=& \frac{qB}{4\mu} K_x y + \frac{q^2 B^2}{8\mu} \rho^2 - \frac{4}{3} \frac{\alpha_s}{r} + \sigma r + ({\bf S}_1\cdot{\bf S}_2) \gamma e^{-\beta r} \, .
\label{eq:pseudopot}
\eqa

\begin{figure}[t!]
\begin{center}
\includegraphics[width=8.1cm]{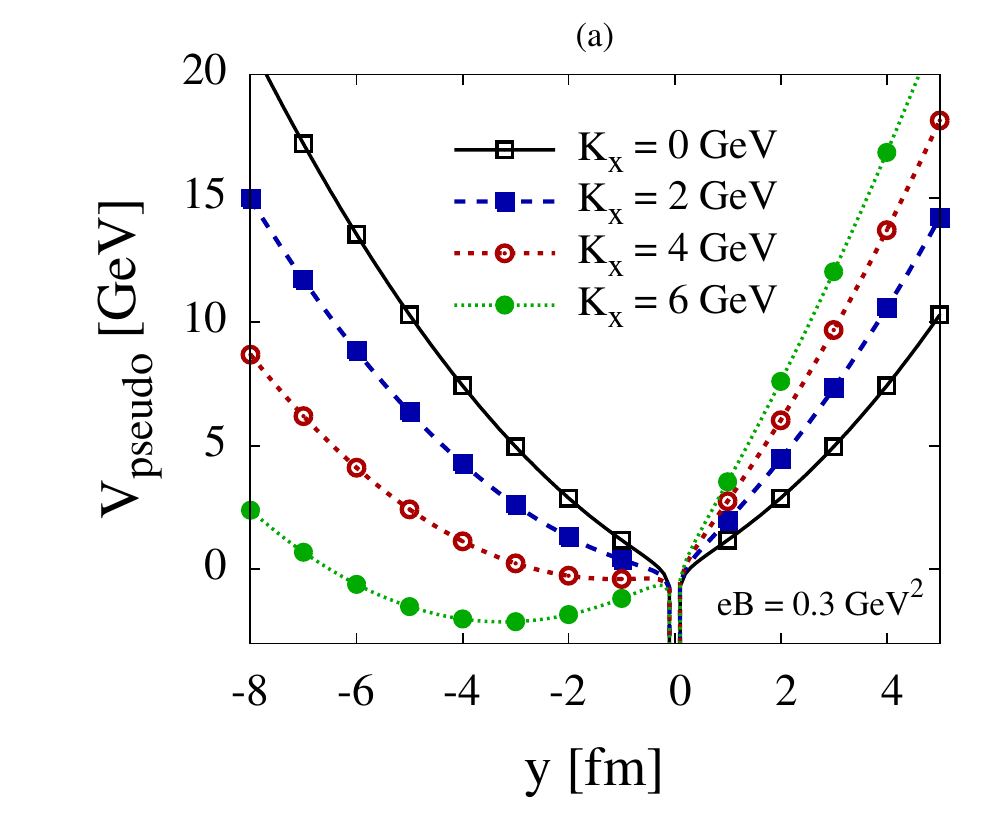}
\includegraphics[width=8.1cm]{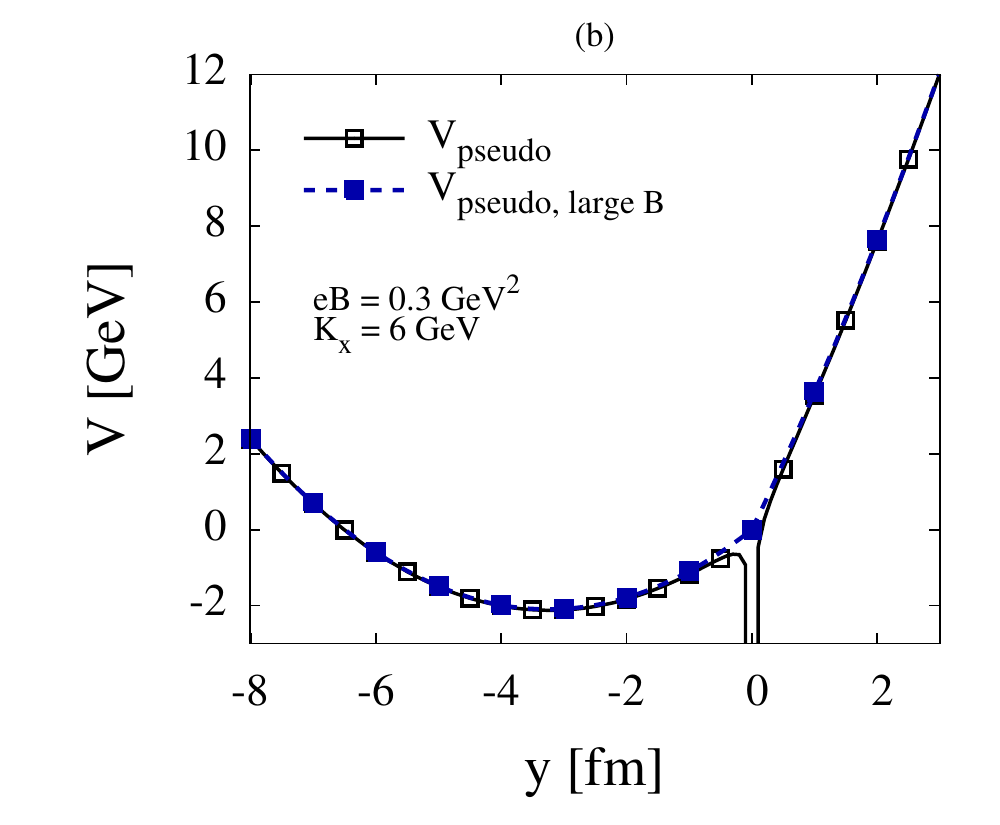}
\end{center}
\vspace{-7mm}
\caption{(a) The pseudopotential (\ref{eq:pseudopot}) as a function of $y$ with $x=z=0$ for charmonium states using the parameters listed in 
App.~\ref{app:tune} Table \ref{tab:charmspec}.  The magnetic field amplitude is assumed to be $eB = 0.3\;{\rm GeV}^2$ and we take $K_x \in \{0,2,4,6\}$ GeV.
(b) Comparison of the exact pseudopotential (\ref{eq:pseudopot}) with the approximate form (\ref{eq:pseudopotlargeB})
for $eB = 0.3\;{\rm GeV}^2$ and $K_x = 6$ GeV.
}
\label{fig:potential1}
\end{figure}

In Fig.~\ref{fig:potential1}(a) we plot the pseudopotential (\ref{eq:pseudopot}) as a function of $y$ with $x=z=0$ for charmonium states using the parameters listed in App.~\ref{app:tune} Table \ref{tab:charmspec}.  The magnetic field amplitude is assumed to be $eB = 0.3\;{\rm GeV}^2$ and we take $K_x \in \{0,2,4,6\}$ GeV.  As can be seen from this figure, at large magnetic field magnitude it is possible for the potential to develop a non-trivial minimum which for positive $K_x$ is at negative $y$.  This minimum is related to so-called motional Stark effect which was originally discussed in \cite{PhysRev.85.259} (see \cite{PhysRevA.4.59} for a discussion in the context of positronium) and recently discussed in the context of quarkonium in Ref.~\cite{Marasinghe:2011bt}.  As a result of this minimum, for large $eB$ and $K_x$ the wave function becomes bilocalized.  For large enough $K_x$ the wavefunction will be dominated by the leftmost minimum and the state will be ``ionized'' by magnetic field; however, we note that this state is, strictly speaking, not a free state since it is still confined in space by the magnetic field.

We note, for later use, that for large $qB$, $K_x$, and $r$, one can ignore the third and fifth terms in (\ref{eq:pseudopot}) to good approximation.  Doing this and setting $x=z=0$ one obtains
\bqa
V_{{\rm pseudo,\;large}\;B}(x=0,y,z=0) &\simeq& \frac{qB}{4\mu} K_x y + \frac{q^2 B^2}{8\mu} y^2 + \sigma |y| \, .
\label{eq:pseudopotlargeB}
\eqa
We compare this approximate form to the exact pseudopotential in Fig.~\ref{fig:potential1}(b) for the case of charm
quarks which have charge $q = 2e/3$.  Based on this 
expression we can find the approximate location of the leftmost minimum
\beq
y_{\rm min} \simeq \frac{4 \sigma \mu - q B K_x}{q^2 B^2} \, ,
\label{eq:y0}
\eeq
from which we learn that for $q B K_x \gtrsim 4 \sigma \mu$ there is a non-trivial minimum at negative $y$.\footnote{If $q$ is negative, the potential minimum appears at positive $y$ instead.}  For charmonium (using the parameters listed in App.~\ref{app:tune} Table \ref{tab:charmspec2}), this translates to the condition
$eB K_x  \gtrsim 0.673 \; {\rm GeV}^3$ and for bottomonium (using the parameters listed in App.~\ref{app:tune} Table \ref{tab:bottomspec}) $eB K_x \gtrsim 5.92 \; {\rm GeV}^3$.
For the maximum magnetic field of $eB = 0.3\;{\rm GeV}^2$ considered herein this translates into the  constraint
$K_x \gtrsim 2.24$ GeV and $K_x \gtrsim 19.7$ GeV for charmonium and bottomonium, respectively. 
For $K_x$ larger 
than these thresholds, the state becomes bilocalized and eventually falls into the ``harmonic'' well.  At this point 
the state is no longer bound by particle-anti-particle interactions, but is instead localized in space by the magnetic field.

\begin{figure}[t!]
\begin{center}
\includegraphics[width=8.15cm]{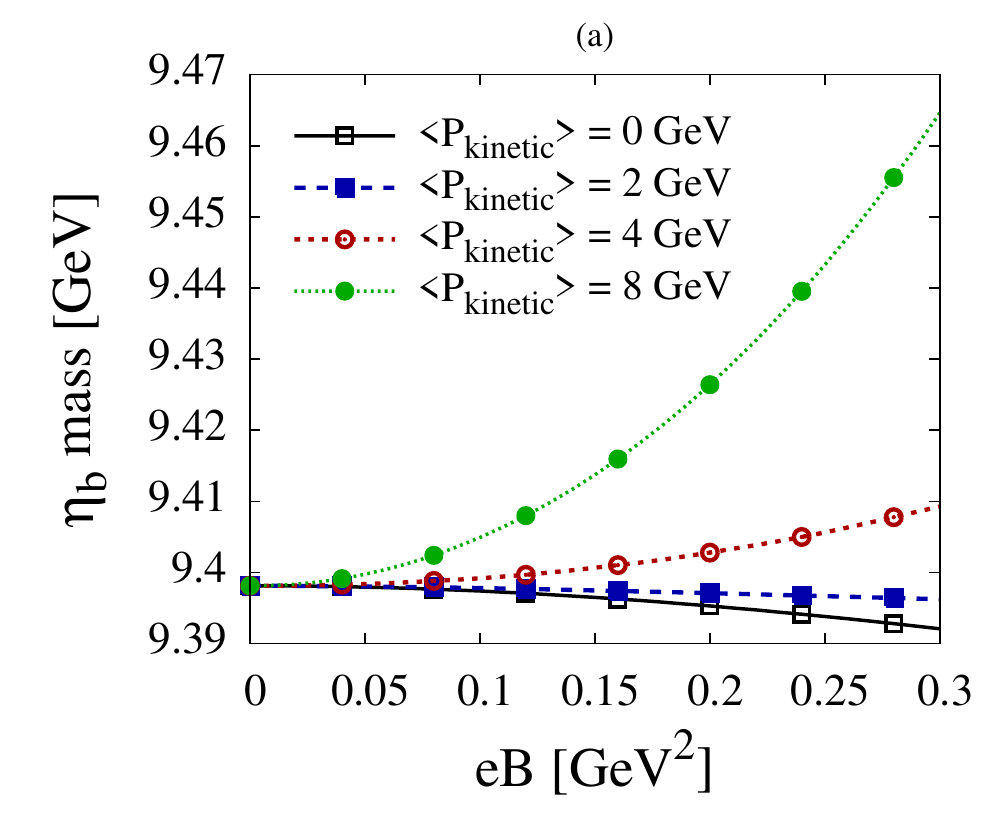}
\includegraphics[width=8.15cm]{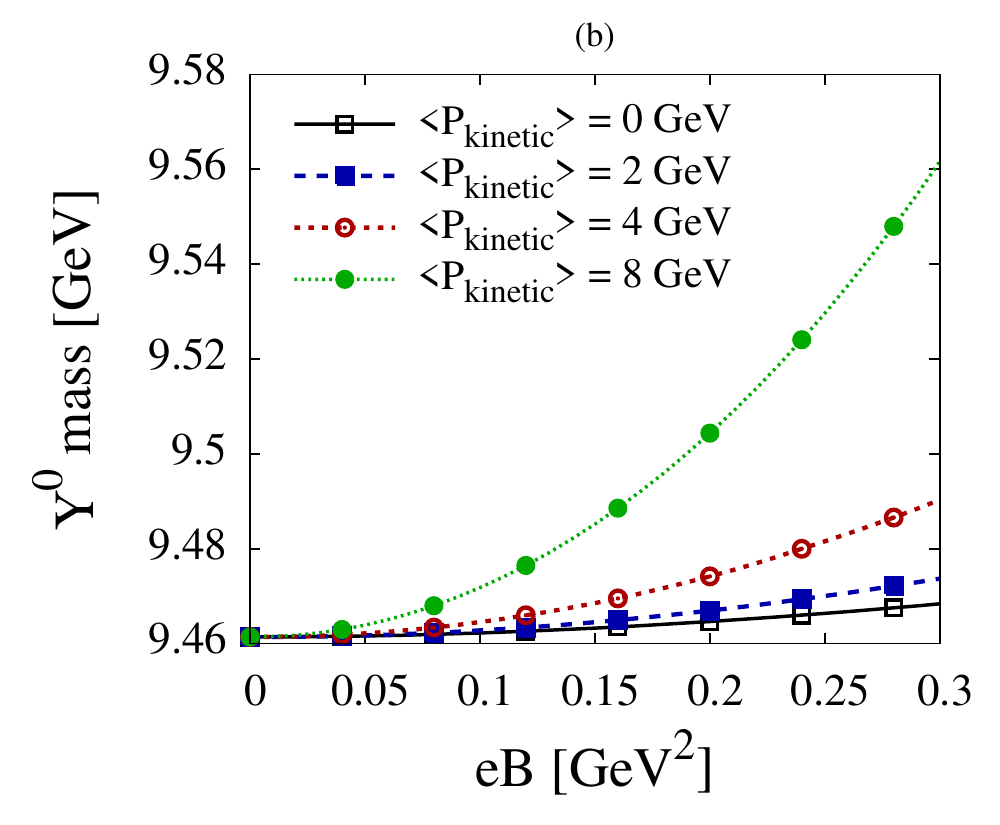}
\includegraphics[width=8.15cm]{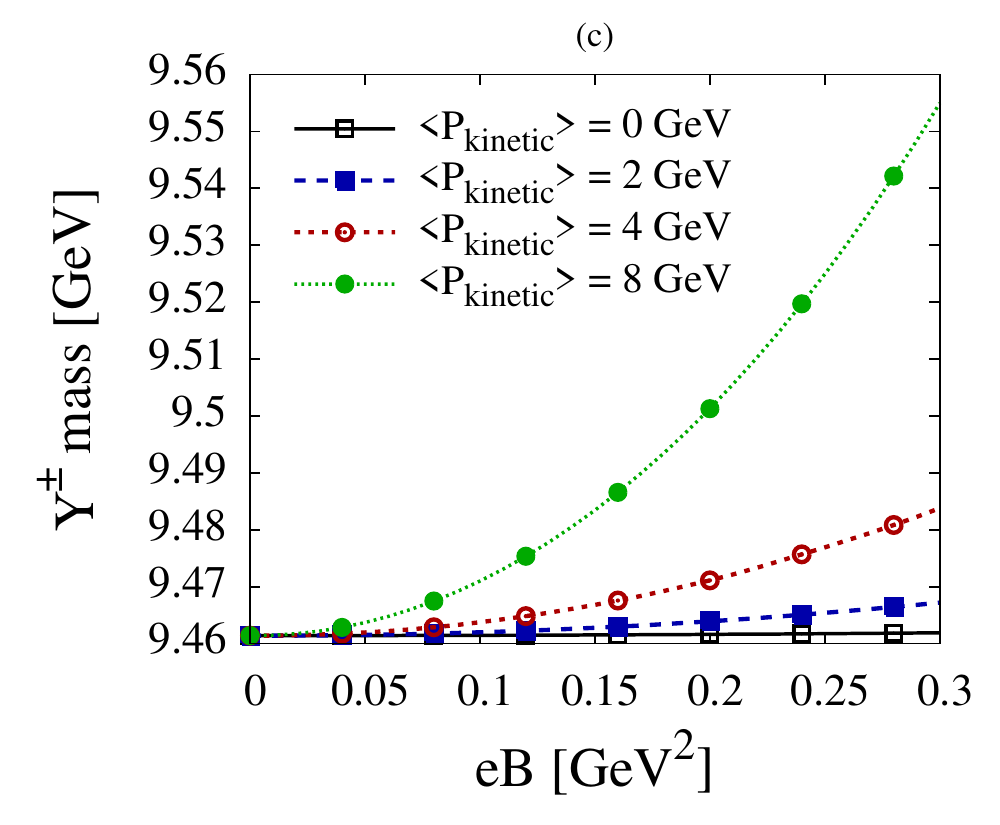}
\end{center}
\vspace{-7mm}
\caption{Masses of the (a) $\eta_b$, (b) $\Upsilon^0$, and (c) $\Upsilon^\pm$ as a function of $eB$ for $\langle P_{\rm kinetic} \rangle \in \{0,2,4,8\}$ GeV.}
\label{fig:bottom1s}
\end{figure}

\begin{figure}[t!]
\begin{center}
\includegraphics[width=9.1cm]{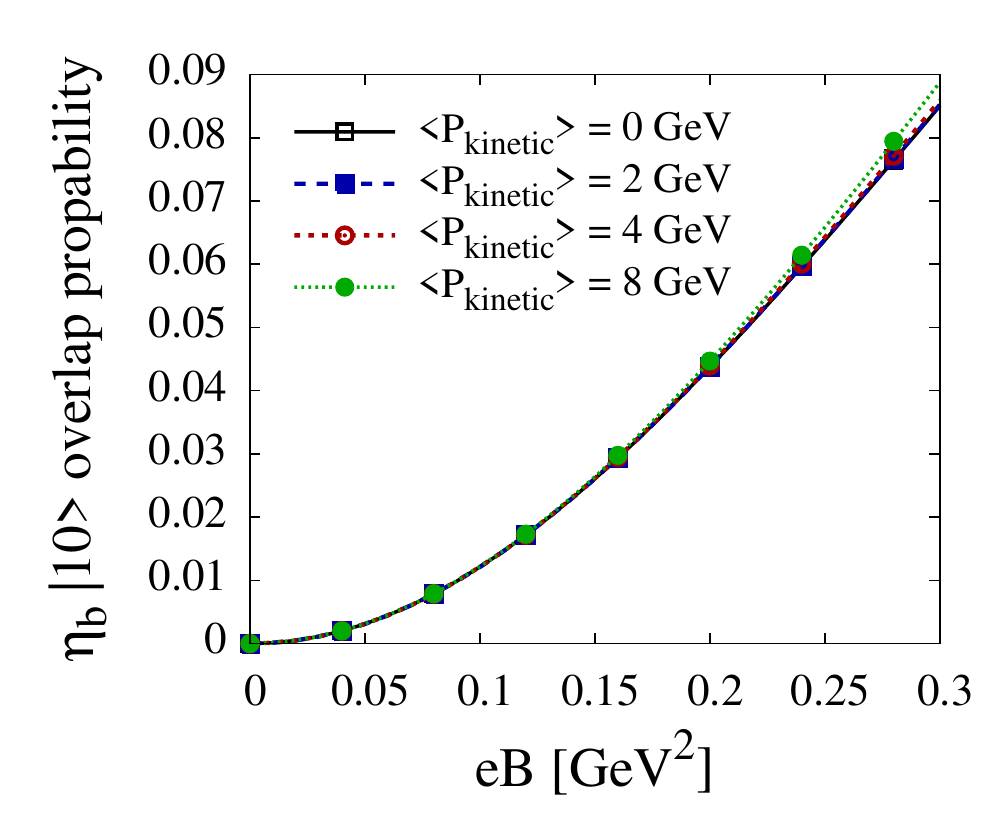}
\end{center}
\vspace{-7mm}
\caption{Probability of finding $|10\rangle$ in the $\eta_b$ state as a function of $eB$ for $\langle P_{\rm kinetic} \rangle \in \{0,2,4,8\}$ GeV. }
\label{fig:etabmix}
\end{figure}

\begin{figure}[t!]
\begin{center}
\includegraphics[width=8.15cm]{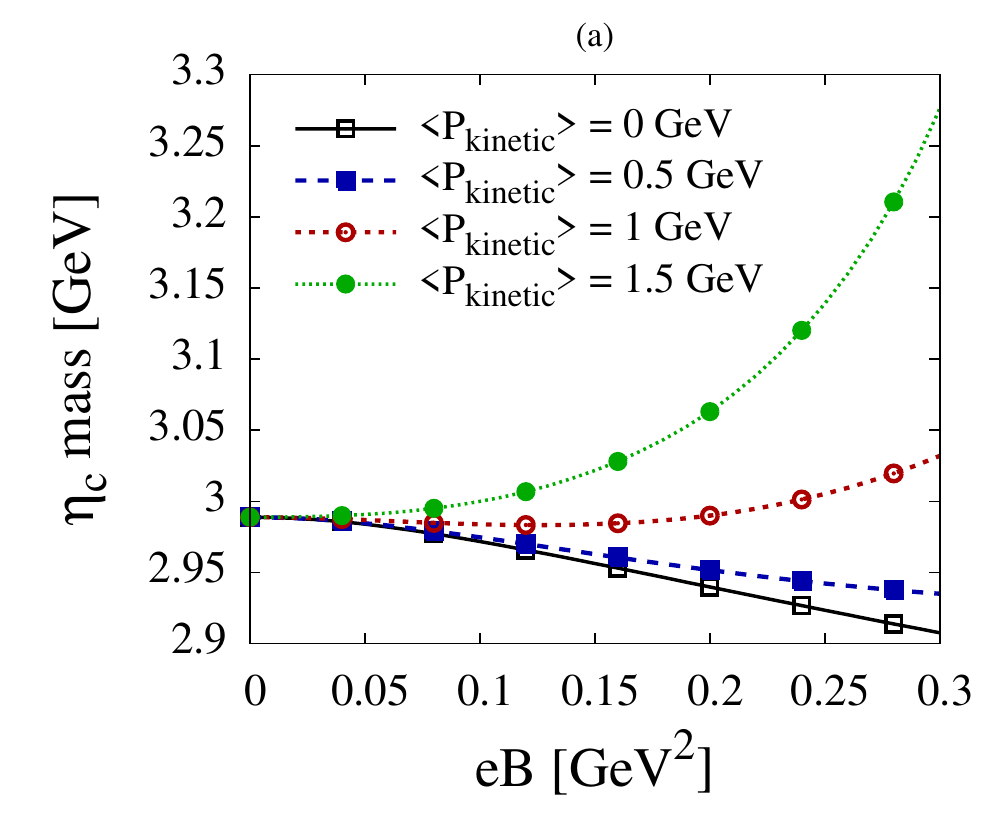}
\includegraphics[width=8.15cm]{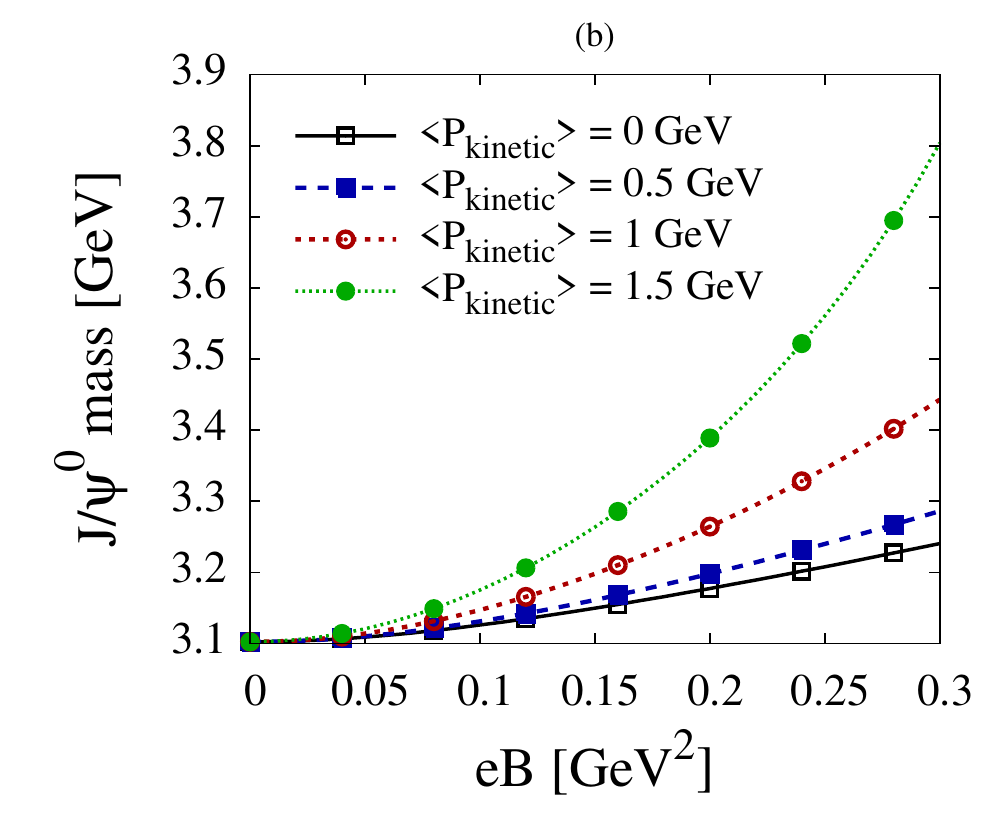}
\includegraphics[width=8.15cm]{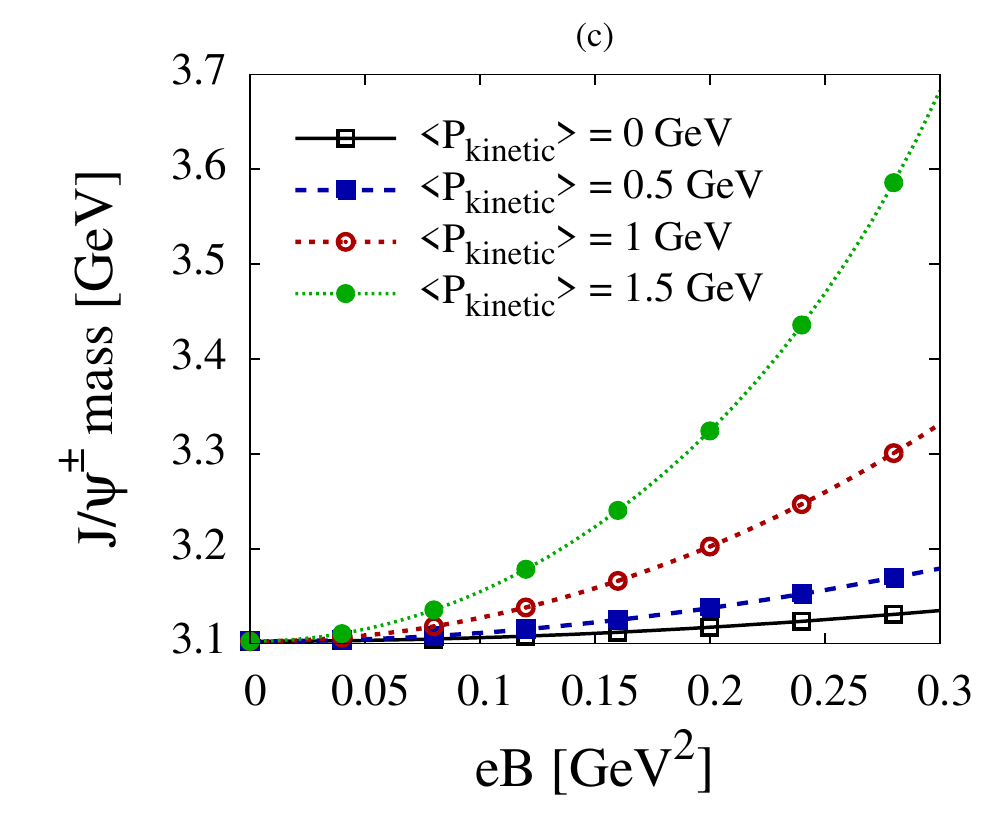}
\end{center}
\vspace{-7mm}
\caption{Masses of the (a) $\eta_c$, (b) $J/\psi^0$, and (c) $J/\psi^\pm$ as a function of $eB$ for $\langle P_{\rm kinetic} \rangle \in \{0,0.5,1,2\}$ GeV.}
\label{fig:jpsi1s}
\end{figure}

\begin{figure}[t!]
\begin{center}
\includegraphics[width=9.1cm]{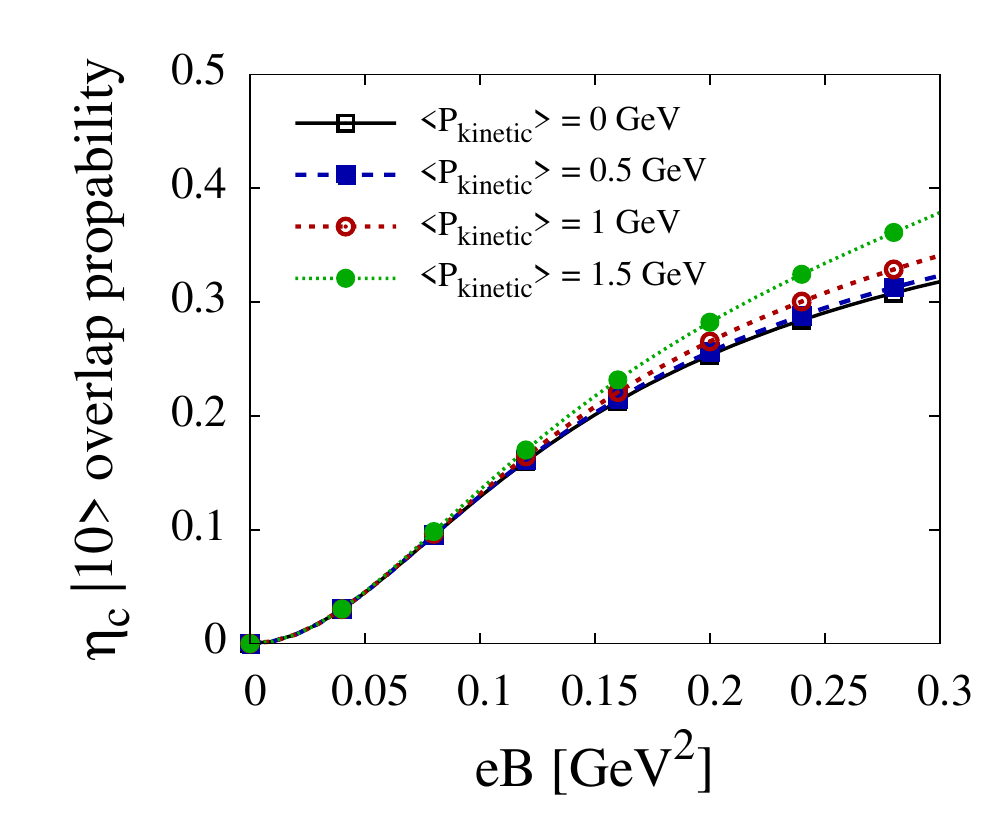}
\end{center}
\vspace{-7mm}
\caption{Probability of finding $|10\rangle$ in the $\eta_c$ state as a function of $eB$ for $\langle P_{\rm kinetic} \rangle \in \{0,0.5,1,2\}$ GeV. }
\label{fig:etacmix}
\end{figure}

In terms of practicalities for the numerics, we note that we use the approximate value in Eq.~(\ref{eq:y0}) to shift the potential along the $y$ direction for large values $K_x$ in order to obtain more accurate numerical results without having to resort to large volumes and/or anisotropic lattices.

\section{Results}
\label{sect:results}

We now present our results using the pseudopotential (\ref{eq:pseudopot}) for both charmonium
and bottomonium states.  For the bottomonium states, the potential parameters and resulting 
vacuum spectra are listed in App.~\ref{app:tune} Table \ref{tab:bottomspec}.  For charmonium
states, the potential parameters and resulting 
vacuum spectra are listed in App.~\ref{app:tune} Table \ref{tab:charmspec2}.  The numerical
algorithm used to find the eigenfunctions and eigenvalues is described in App.~\ref{app:nummethod}.
We note that we have tested the numerical algorithm using a harmonic interaction and have 
found agreement between the extracted wave functions, energy eigenvalues, etc. and the analytic
formulae presented in previous sections to within machine precision.  This gives us confidence
in our numerical method.

\subsection{Bottomonia}

We first consider bottomium states.  In Fig.~\ref{fig:bottom1s} we plot the masses of the (a) $\eta_b$, (b) $\Upsilon^0$, and (c) $\Upsilon^\pm$ as a function of $eB$ for $\langle P_{\rm kinetic} \rangle \in \{0,2,4,8\}$ GeV.  For $\langle P_{\rm kinetic} \rangle = 0$ GeV, we see the pattern expected, namely that the $\eta_b$ mass is lowered due to spin-mixing, the $\Upsilon^0$ mass increases for the same reason, and the $\Upsilon^\pm$ states are very-weakly affected (there is a small change in the mass due to the magnetic potential effects, but it is negligible).  As we increase $\langle P_{\rm kinetic} \rangle$, we see that the masses of all states increase.  The result is in agreement with what we obtained analytically for the harmonic interaction (see first term in Eq.~(\ref{eq:harmensub})).  For $\langle P_{\rm kinetic} \rangle=0$ and $eB = 0.3\;{\rm GeV}^2$ one sees a 0.06\% decrease in the mass of the $\eta_b$.  For $\langle P_{\rm kinetic} \rangle=8\;{\rm GeV}$, one sees an increase of 0.71\% in the $\eta_b$ mass.  For the $\Upsilon$ states, the mass is a monotonically increasing function of $eB$ and $\langle P_{\rm kinetic} \rangle$.  The maximum mass increase is on the order of 1.1\% for the $\Upsilon$ states.

Based on the findings above one can estimate the effect of strong magnetic fields on bottomonium production in the LHC heavy ion collisions ($eB \sim 0.3\;{\rm GeV}^2$).  The cross sections for quarkonium production from both gluon-gluon fusion and quark-antiquark annihilation both scale (to leading order) as $M^{-2}$.  Assuming that we need only build in the mass correction in order
to account for the magnetic field, the maximal effect on 1s bottomonium states can be estimated to be on the order of a 2\% effect.

We can extract the energy difference between the singlet and triplet states to determine the overlap probability for the $|10\rangle$ (triplet) state with, e.g. the $\eta_b$ state, via Eq.~(\ref{eq:spineigenstates}).  In vacuum, the $\eta_b$ is a pure singlet state, however, a background magnetic field causes a mixing of the singlet and triplet states.  In Fig.~\ref{fig:etabmix} we
plot the $\eta_b$ triplet overlap probability as a function of $eB$ for $\langle P_{\rm kinetic} \rangle \in \{0,2,4,8\}$ GeV.  As we can see from this figure, at LHC energies one estimates the overlap probability to be approximately 8.5\%.  This percentage of $\eta_b$ states would be able
to decay through dilepton decay.  Correspondingly, there would be an 8.5\% reduction in the dilepton
decays from the $\Upsilon^0$ state.  The $\Upsilon^\pm$ states do not mix and would not have their dilepton decay rate modified.  Averaging over the three different types of $\Upsilon$ states we would predict an approximately 2.8\% suppression of $\Upsilon(1s)$ decays.  The dileptons which failed to come from the $\Upsilon^0$ decays, would instead appear at the mass of the $\eta_b$ state.  This would manifest itself through a peak in the dilepton spectrum at the $\eta_b$ invariant mass. We note, however, that given finite detector resolution, it may not be possible to experimentally resolve this feature in the dilepton spectrum invariant mass spectrum.  The splitting between the $\eta_b$ and $\Upsilon$ vacuum masses is approximately 63 MeV and this is only weakly dependent on the magnetic field.  The CMS and ALICE experiments have a invariant mass resolution on the order of 100 MeV \cite{CMSres,ALICEres} so they would not be able to see this effect, instead they would see a slight broadening of the $\Upsilon(1s)$ peak.

\subsection{Charmonia}

We now turn our attention to the charmonium states.  In Fig.~\ref{fig:jpsi1s} we plot the masses of the (a) $\eta_c$, (b) $J/\psi^0$, and (c) $J/\psi^\pm$ as a function of $eB$ for $\langle P_{\rm kinetic} \rangle \in \{0,0.5,1,1.5\}$ GeV using the charmonium-tuned parameters listed in App.~\ref{app:tune} Table \ref{tab:charmspec2}.~\footnote{
For a comparison of the results obtained using the bottom-tuned potential applied to charmonium states, see
Fig.~\ref{fig:jpsi1s-comp} in App.~\ref{app:tune} and the surrounding discussion.}
  
For $\langle P_{\rm kinetic} \rangle = 0$ GeV, we see the pattern expected, namely that the $\eta_c$ mass is lowered due to spin-mixing, the $J/\psi^0$ mass increases for the same reason, and the $J/\psi^\pm$ states are weakly affected.  For $K_x=0$ and $eB = 0.3\;{\rm GeV}^2$ one sees a 3.5\% decrease in the mass of the $\eta_c$.  For $K_x=1.5\;{\rm GeV}$, one sees an increase of 19\% in the $\eta_c$ mass.  For the $J/\psi$ states, the mass is a monotonically increasing function of $eB$ and $\langle P_{\rm kinetic} \rangle$.  The maximum mass increase is on the order of 31\% for the $J/\psi$ states.  Again assuming that to leading order the $J/\psi$ production cross section scales like $M^{-2}$ one can estimate that this would result in a maximum suppression of $J/\psi$ my approximately 42\%, with the corresponding nuclear suppression being $R_{AA} \sim 0.58$.

In Fig.~\ref{fig:etacmix} we
plot the $\eta_c$ triplet overlap probability as a function of $eB$ for $\langle P_{\rm kinetic} \rangle \in \{0,0.5,1,1.5\}$ GeV.  As we can see from this figure, at LHC energies one estimates the overlap probability to be approximately 32\%.  This percentage of $\eta_c$ states would be able
to decay through dilepton decay.  Correspondingly, there would be a 32\% reduction in the dilepton
decays from the $J/\psi^0$ state.  The $J/\psi^\pm$ states do not mix and would not have their dilepton decay rate modified.  Averaging over the three different types of $J/\psi$ states we would predict an approximately 11\% suppression of $J/\psi$ decays.  The dileptons which failed to come from the $J/\psi^0$ decays, would instead appear at the mass of the $\eta_c$ state.  This would manifest itself through a peak in the dilepton spectrum at the $\eta_c$ invariant mass.  Regarding the feasibility of measuring this effect experimentally, the splitting between the $\eta_c$ and $J/\psi$ vacuum masses is approximately 113 MeV and the CMS and ALICE experiments have a invariant mass resolution on the order of 30 MeV \cite{CMSres,ALICEres}.  As a result, it may be possible see hints of this effect in the charmonium sector.  To truly confirm this effect, however, it would seem that either the detector resolution or the 
Crystal Ball function would need to be improved upon.

\section{Conclusions}
\label{sect:conc}

In this paper we have made a first investigation of the effects of an external magnetic
field on charmonium and bottomonium states.  We have taken into account the external
potential associated with the magnetic field, motional effects, and the singlet-triplet
mixing of states.  We solved the resulting three-dimensional Schr\"odinger equation
analytically for the case of a harmonic interaction and numerically for a realistic
quarkonium potential consisting of a Cornell potential plus a spin-spin interaction.
We demonstrated that it is not possible to fully factorize the Hamiltonian of the two-particle
system in the presence of the magnetic field.  Instead, one can introduce a conserved 
quantity called the pseudomomentum, ${\bf K}$, which allows one to write a compact 
``pseudopotential'' for the system which has a non-trivial dependence on the components
of ${\bf K}$ that are perpendicular to the magnetic field.  We then derived a general relation 
between the pseudomomentum and the kinetic COM momentum of the system.  For the
harmonic interaction, the latter relation could be derived analytically for all states.

Herein we have considered states with COM momentum up to 1.5 GeV in the case of 1s $J/\psi$ 
and 10 GeV in the case of the $\Upsilon(1s)$.  For $J/\psi$ COM momentum
larger than this threshold we find that the state will dissociate in the magnetic field
(a similar conclusion but with a different threshold was found in Ref.~\cite{Marasinghe:2011bt});
however, since our results were derived in the context of a non-relativistic limit, one
expects relativistic corrections to become quantitatively important at large momenta.  For this reason,
it seems necessary to reformulate the problem in a relativistic framework if one
wants to arrive at more reliable conclusions about the phenomenological consequences on 
$J/\psi$ production.  For $\Upsilon$ production, the threshold for magnetic field 
dissociation is estimated to be on the order of 20 GeV.  At these high momenta, a relativistic treatment of 
the COM motion is necessary; however, for the range of $\Upsilon$ COM momenta considered herein a non-relativistic
treatment should be reasonable.  Our results indicate that the maximal effect 
on $\Upsilon$ production is on the order of 2\% and, as a result, it is probably safe to 
ignore this effect on theses states.  For both systems, in order to minimize the effect of magnetic fields in experimental measurements of quarkonium suppression, one can apply transverse momentum cuts which eliminate states
with high COM momentum.

As part of the analysis we presented a quantitative analysis of the effect of singlet-triplet spin
mixing for both charmonium and bottomonium 1s states.  The effect causes an increase in
the mass of the $|10\rangle$ triplet state and a decrease in the mass of the $|00\rangle$
state.  In addition, because of the mixing, some decays of the $|10\rangle$ will appear
instead at the mass of the $|00\rangle$ state; however, given the fact that the splittings
in the charmonium and bottomonium states are on the order of 113 and 62 MeV, respectively,
it does not seem possible to use existing experimental configurations to fully resolve this
effect.  With limited resolution, the mixing would appear instead as a broadening of the triplet
state peak.

The estimates of the phenomenological effect of static magnetic fields obtained herein 
are subject to two important caveats: (1) our investigations were restricted to the vacuum 
Cornell potential plus a spin-spin interaction and (2) we did not investigate the effect
on excited states.  Regarding caveat number one, in a future study we plan to include 
finite-temperature effects on the potential (see e.g. \cite{Strickland:2011mw,Strickland:2011aa}) 
and to simultaneously include more realistic vacuum potentials (see e.g. 
\cite{Haglin:2002ga,Barnes:2005pb,Radford:2007vd,Repko:2012rk,Laschka:2012cf}).  Since 
finite temperature effects reduce the binding energy and cause the states to be more extended 
in space, one can expect a priori that the magnetic field effect will be larger at finite 
temperature.  Regarding caveat number two, we also plan a thorough investigation of magnetic field
effects on excited states using realistic potential models.  
The effects on excited states are expected to be more important
than on the ground state for two reasons: (a) excited states are more extended in space and are
therefore more sensitive to the quadratic magnetic potential and (b) spin-mixing effects
grow larger as the angular momentum representation of the state increases.  Since 
excited state feed-down makes up on the order of 50\% of both $J/\psi$ and $\Upsilon$
production one expects this to affect the ground states themselves.

Based on the two caveats laid out in the preceding paragraph, we expect that our estimates 
of the effect of static magnetic fields on heavy quarkonium production are a lower bound.
That being said, one should also take into account the fact that the magnetic field
generated in a heavy ion collision is neither static nor constant in space.  One expects
very strong magnetic fields only for the first 1-2 fm/c after the initial nuclear impact
and as a result this would act to reduce the integrated magnetic field effect.  
In addition, it will be necessary to make a detailed investigation
of the effect of magnetic field on the string tension and finite-temperature screened potential.
We plan to investigate these effects in a future study.  
In closing, we have demonstrated in this paper that the effect of magnetic fields on heavy 
quarkonium, particularly the $J/\psi$, warrants further investigation.
We have laid the ground work for such studies in the paper.

\section*{Acknowledgements}

We thank F.S. Navarra and J. Noronha for motivation and useful discussions.  J.A. was
supported by DOE Grant No. DE-FG02-89ER40531.  M.S. was supported in part by DOE Grant No. 
DE-SC0004104.


\appendix

\section{Potential Tuning}
\label{app:tune}

In this appendix we present comparisons of bottomonium and charmonium state masses computed
using the model potential (\ref{eq:potmodel}) and experimental data \cite{Beringer:1900zz}.
We present results from the two different ``tunings'' which are used in the body of the 
manuscript separately.

\subsection{Bottom-tuned potential}

\begin{table}[t]
\begin{center}
\setlength{\tabcolsep}{0.4em}
\begin{tabular}{|c|c|c|c|c|}
\hline
{\bf \small State} & {\bf \small Name} &  {\bf \small Exp.~\cite{Beringer:1900zz}} & {\bf \small Model} & {\bf Rel.~Err.} \\ 
\hline
$1^1S_0$ & $\eta_b(1S)$  &  9.398 {\rm GeV} & 9.398 {\rm GeV} & 0.001\% \\ 
\hline
$1^3S_1$ & $\Upsilon(1S)$  &  9.461 {\rm GeV} & 9.461 {\rm GeV} & 0.004\% \\ 
\hline
$1^3P_0$ & $\chi_{b0}(1P)$  &  9.859 {\rm GeV} & \multirow{4}{*}{9.869 {\rm GeV}} & \multirow{4}{*}{0.21\%} \\ 
\cline{1-3}
$1^3P_1$ & $\chi_{b1}(1P)$  &  9.893 {\rm GeV} & & \\ 
\cline{1-3}
$1^3P_2$ & $\chi_{b2}(1P)$  &  9.912 {\rm GeV} & & \\ 
\cline{1-3}
$1^1P_1$ & $h_b(1P)$  &  9.899 {\rm GeV} & & \\ 
\hline
$2^1S_0$ & $\eta_b(2S)$  &  9.999 {\rm GeV} & 9.977 {\rm GeV} & 0.22\% \\ 
\hline
$2^3S_1$ & $\Upsilon(2S)$  &  10.002 {\rm GeV} & 9.999 {\rm GeV} & 0.03\% \\ 
\hline
$2^3P_0$ & $\chi_{b0}(2P)$  &  10.232 {\rm GeV} & \multirow{4}{*}{10.246 {\rm GeV}} & \multirow{4}{*}{0.05\%} \\ 
\cline{1-3}
$2^3P_1$ & $\chi_{b1}(2P)$  &  10.255 {\rm GeV} & & \\ 
\cline{1-3}
$2^3P_2$ & $\chi_{b2}(2P)$  &  10.269 {\rm GeV} & & \\ 
\cline{1-3}
$2^1P_1$ & $h_b(2P)$  & - & & \\ 
\hline
$3^1S_0$ & $\eta_b(3S)$  &  - & 10.344 {\rm GeV} & - \\ 
\hline
$3^3S_1$ & $\Upsilon(3S)$  &  10.355 {\rm GeV} & 10.358 {\rm GeV} & 0.03\% \\ 
\hline
\end{tabular}
\end{center}
\caption{Comparison of experimentally measured particle masses from Ref.~\cite{Beringer:1900zz} for the bottomonium
system with ``bottom-tuned'' model predictions obtained using the potential model specified in Eq.~(\ref{eq:potmodel}).
The parameters used were $m_b = 4.7~{\rm GeV}$, 
$\gamma = 0.318~{\rm GeV}$, $\beta = 1.982~{\rm GeV}$, $\alpha_s = 0.315443$, and $\sigma = 0.210~{\rm GeV}^2$.
In the case that there is no experimental data, we indicate this with a dash.
}
\label{tab:bottomspec}
\end{table}

In Table \ref{tab:bottomspec} we compare bottomonia experimental data and the ``bottom-tuned'' potential model. 
The model results were computed on a lattice size of $256^3$ with lattice spacing of $a=0.1~{\rm GeV}^{-1}$.  
The parameters used were $m_b = 4.7~{\rm GeV}$, 
$\gamma = 0.318~{\rm GeV}$, $\beta = 1.982~{\rm GeV}$, $\alpha_s = 0.315443$, and $\sigma = 0.210~{\rm GeV}^2$.
Note that, since the potential model used herein does not include spin-orbit or tensor interactions, the model does not 
predict a splitting between the $\chi$ states.  For these states, the error reported is computed from the average of the experimental masses.

In Table \ref{tab:charmspec} we compare charmonia experimental data and the ``bottom-tuned'' potential model. 
The model results were computed on a lattice size of $256^3$ with lattice spacing of $a=0.2~{\rm GeV}^{-1}$.  
The parameters used were $m_c = 1.29~{\rm GeV}$, 
$\gamma = 0.825~{\rm GeV}$, $\beta = 1.982~{\rm GeV}$, $\alpha_s = 0.315443$, and $\sigma = 0.210~{\rm GeV}^2$.

\begin{table}[t]
\begin{center}
\setlength{\tabcolsep}{0.4em}
\begin{tabular}{|c|c|c|c|c|}
\hline
{\bf \small State} & {\bf \small Name} &  {\bf \small Exp.~\cite{Beringer:1900zz}} & {\bf \small Model} & {\bf Rel.~Error} \\ 
\hline
$1^1S_0$ & $\eta_c(1S)$  &  2.984 {\rm GeV} & 3.048 {\rm GeV} & 2.2\% \\ 
\hline
$1^3S_1$ & $J/\psi(1S)$  &  3.097 {\rm GeV} & 3.100 {\rm GeV} & 0.11\% \\ 
\hline
$2^1S_0$ & $\eta_c(2S)$  &  3.639 {\rm GeV} & 3.721 {\rm GeV} & 2.3\% \\ 
\hline
$2^3S_1$ & $J/\psi(2S)$  &  3.686 {\rm GeV} & 3.748 {\rm GeV} & 1.7\% \\ 
\hline
\end{tabular}
\end{center}
\caption{Comparison of experimentally measured particle masses from Ref.~\cite{Beringer:1900zz} for the charmonium
system with ``bottom-tuned'' model predictions obtained using the potential model specified in Eq.~(\ref{eq:potmodel}).
The parameters used were $m_c = 1.29~{\rm GeV}$, 
$\gamma = 0.825~{\rm GeV}$, $\beta = 1.982~{\rm GeV}$, $\alpha_s = 0.315443$, and $\sigma = 0.210~{\rm GeV}^2$.}
\label{tab:charmspec}
\end{table}

\subsection{Charm-tuned potential}
\label{subsect:charmtuned}

\begin{table}[t!]
\begin{center}
\setlength{\tabcolsep}{0.4em}
\begin{tabular}{|c|c|c|c|c|}
\hline
{\bf \small State} & {\bf \small Name} &  {\bf \small Exp.~\cite{Beringer:1900zz}} & {\bf \small Model} & {\bf Rel.~Error} \\
\hline
$1^1S_0$ & $\eta_c(1S)$  &  2.984 {\rm GeV} & 2.989 {\rm GeV} & 0.16\% \\ 
\hline
$1^3S_1$ & $J/\psi(1S)$  &  3.097 {\rm GeV} & 3.102 {\rm GeV} & 0.17\% \\ 
\hline
$2^1S_0$ & $\eta_c(2S)$  &  3.639 {\rm GeV} & 3.590 {\rm GeV} & 1.3\% \\ 
\hline
$2^3S_1$ & $J/\psi(2S)$  &  3.686 {\rm GeV} & 3.650 {\rm GeV} & 0.97\% \\ 
\hline
\end{tabular}
\end{center}
\caption{Comparison of experimentally measured particle masses from Ref.~\cite{Beringer:1900zz} for the charmonium
system with ``charm-tuned'' model predictions obtained using the potential model specified in Eq.~(\ref{eq:potmodel}).
The parameters used were $m_c = 1.29~{\rm GeV}$, 
$\gamma = 2.06~{\rm GeV}$, $\beta = 1.982~{\rm GeV}$, $\alpha_s = 0.234$, and $\sigma = 0.174~{\rm GeV}^2$.}
\label{tab:charmspec2}
\end{table}

\begin{figure*}[t!]
\begin{center}
\includegraphics[width=8.1cm]{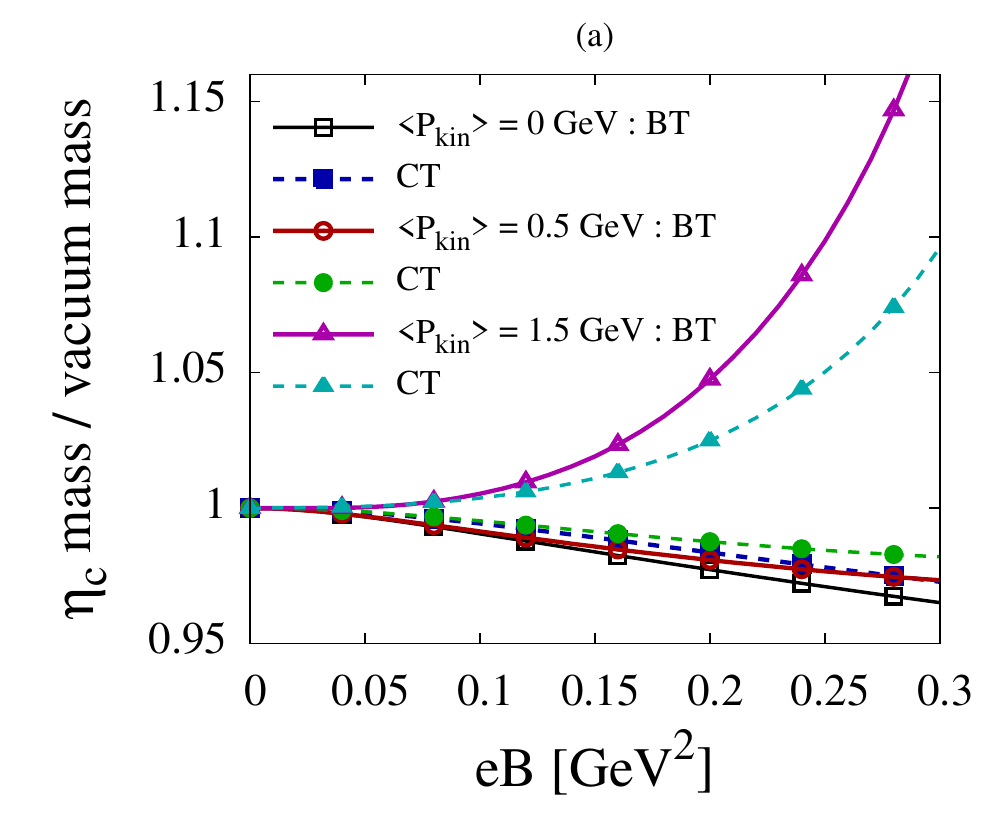}
\includegraphics[width=8.1cm]{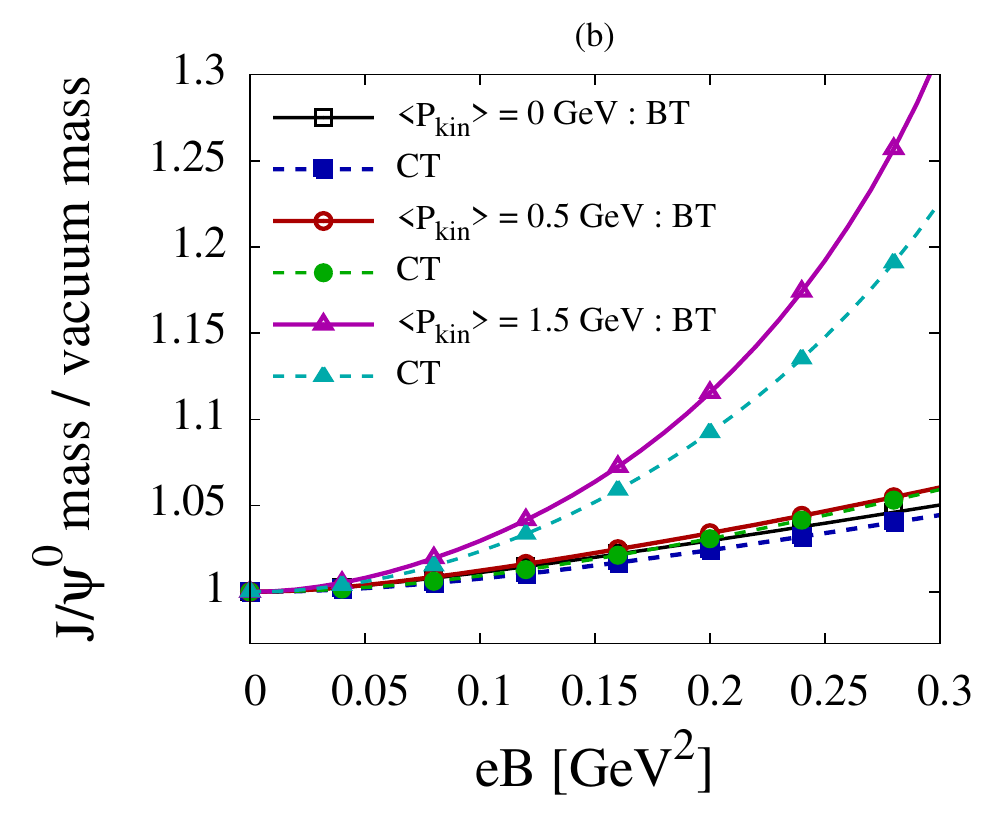}\\
\includegraphics[width=8.1cm]{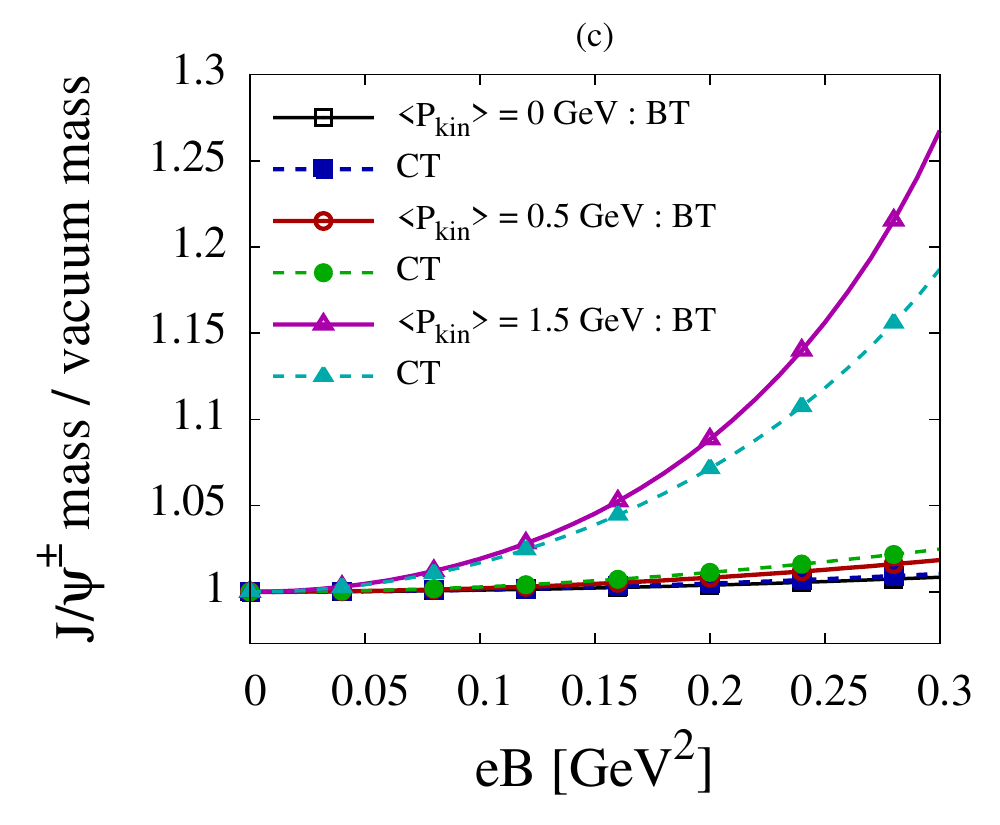}
\includegraphics[width=8.1cm]{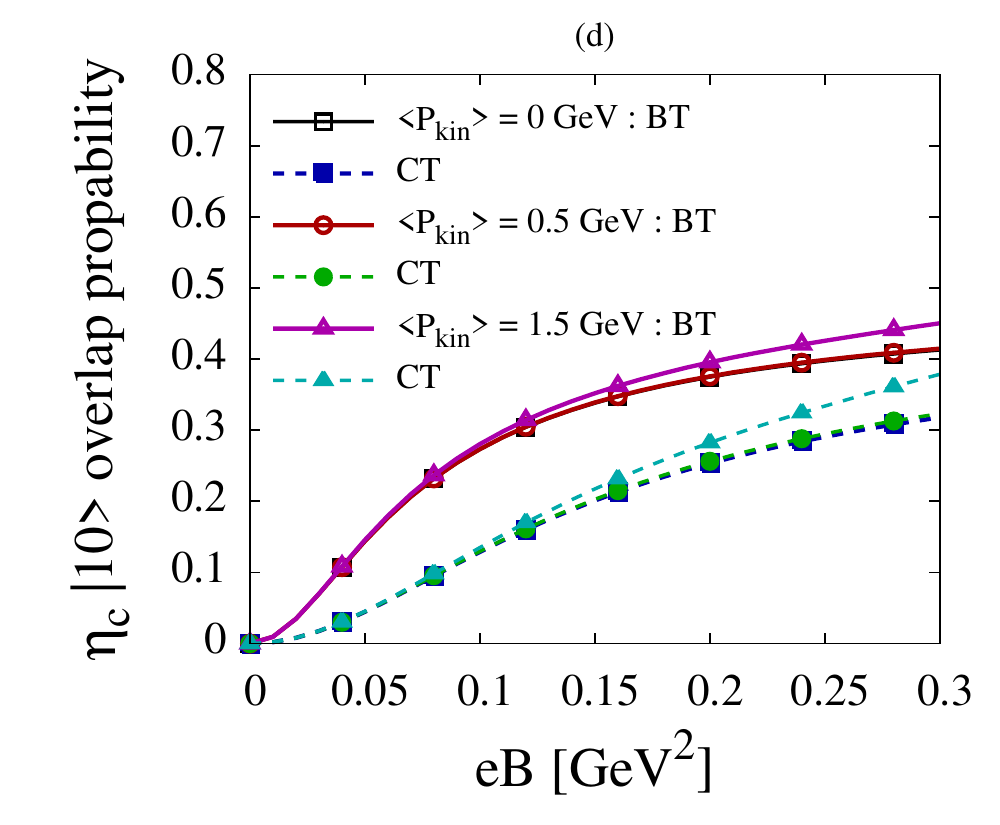}
\end{center}
\vspace{-7mm}
\caption{Comparison of (a) $\eta_c$, (b) $J/\psi^0$, and (c) $J/\psi^\pm$ masses divided by the $eB=0$ vacuum masses and (d) triplet overlap probability as a function of $eB$ for $\langle P_{\rm kinetic} \rangle \in \{0,0.5,1,1.5\}$ GeV.  BT and CT indicate the results obtained using the bottom-tuned (Table \ref{tab:charmspec}) and charm-tuned (Table \ref{tab:charmspec2}) potentials, respectively.}
\label{fig:jpsi1s-comp}
\end{figure*}

In Table \ref{tab:charmspec} we present a second parameter tuning which better reproduces the energy levels of low-lying charmonium states. 
As can be seen from this table, even when tuned to the charmonium states, the relative errors of the heavy quark potential model spectra compared to experimental data are larger than those obtained for bottomonium states.  This is to be expected and indicates that it is necessary to include relativistic corrections to obtain a more accurate reproduction of the spectrum of charmonium states.  Comparing the relative errors of charmonia masses using the bottom-tuned and charm-tuned potential we expect that the charm-tuned potential is a better approximation than the bottom-tuned potential since the singlet-triplet split is very close to the experimentally determined splitting.  That being said, we can use those two tunings to assess the dependence of our results on the assumed quark interaction potential.  In Fig.~\ref{fig:jpsi1s-comp} we show the scaled masses and triplet overlap probabilities using the two different tunings.  In the figure, the bottom-tuned results are indicated by ``BT'' and the charm-tuned results by ``CT''.  As we can see from this figure, the results obtained with the two different tunings are in qualitative agreement; however, we reiterate that we expect the charm-tuned results to be a better approximation.

\section{Numerical Method}
\label{app:nummethod}

To solve the resulting Schr\"odinger equation we
use the finite difference time domain method~\cite{Sudiarta:2007,Strickland:2009ft,Margotta:2011ta}.  
Here we briefly review the technique.  
To determine the wave functions of bound quarkonium states, we must solve
the time-independent Schr\"odinger equation for the relative wave function
\bqa
\hat{H}_{\rm rel} \Psi_\upsilon({\bf r}) &=& E_\upsilon \, \Psi_\upsilon({\bf r}) \, , 
\label{3dSchrodingerEQ}
\eqa
on a three-dimensional lattice in coordinate space.  
The index $\upsilon$ on the eigenfunctions,
$\phi_\upsilon$, and energies, $E_\upsilon$, represents a list of all
relevant quantum numbers.  To obtain the time-independent eigenfunctions 
we start with the time-dependent Schr\"odinger equation
\beq
i \frac{\partial}{\partial t} \Psi({\bf x},t) = \hat{H}_{\rm rel} \Psi({\bf x},t) \, ,
\label{3dSchrodingerEQminkowski}
\eeq
which can be solved by expanding in terms of the eigenfunctions,
$\Psi_\upsilon({\bf r})$:
\beq \Psi({\bf r},t) = \sum_\upsilon c_\upsilon \Psi_\upsilon({\bf r})
e^{- i E_\upsilon t}~.
\label{eigenfunctionExpansionMinkowski}
\eeq
If one is only interested in the lowest energy states (ground state
and first few excited states) an efficient way to proceed is to
transform~(\ref{3dSchrodingerEQminkowski})
and~(\ref{eigenfunctionExpansionMinkowski}) to Euclidean time using a
Wick rotation, $\tau \equiv i t$:
\beq \frac{\partial}{\partial \tau} \Psi({\bf r},\tau) = - \hat{H}_{\rm rel}
\psi({\bf r},\tau) \, ,
\label{3dSchrodingerEQeuclidean}
\eeq
and
\beq \Psi({\bf r},\tau) = \sum_\upsilon c_\upsilon \Psi_\upsilon({\bf
r}) e^{- E_\upsilon \tau} ~.
\label{eigenfunctionExpansionEuclidean}
\eeq
For details of the discretizations used etc. we refer the reader 
to Refs.~\cite{Strickland:2009ft}.

\subsection{Finding the ground state}

By definition, the ground state is the state with the lowest energy
eigenvalue, $E_0$. Therefore, at late imaginary time the sum
over eigenfunctions (\ref{eigenfunctionExpansionEuclidean}) is
dominated by the ground state eigenfunction
\beq \lim_{\tau \rightarrow \infty} \Psi({\bf r},\tau) \rightarrow c_0
\Psi_0({\bf r}) e^{- E_0 \tau}~.
\label{groundstateEuclideanLateTime}
\eeq
Due to this, one can obtain the ground state wavefunction,
$\phi_0$, and energy, $E_0$, by solving
Eq.~(\ref{3dSchrodingerEQeuclidean}) starting from a random
three-dimensional wavefunction, $\Psi_{\text{initial}}({\bf r},0)$,
and evolving forward in imaginary time. The initial wavefunction
should have a nonzero overlap with all eigenfunctions of the
Hamiltonian; however, due to the damping of higher-energy
eigenfunctions at sufficiently late imaginary times we are left with
only the ground state, $\Psi_0({\bf r})$. Once the ground state
wavefunction (or any other wavefunction) is found, we can
compute its energy eigenvalue via
\bqa
E_\upsilon(\tau\to\infty) = \frac{\langle \Psi_\upsilon | \hat{H} |
\Psi_\upsilon \rangle}{\langle \Psi_\upsilon | \Psi_\upsilon
\rangle} = \frac{\int d^3{\bf x} \, \Psi_\upsilon^*
\, \hat{H} \, \Psi_\upsilon }{\int d^3{\bf x} \, \Psi_\upsilon^*
\Psi_\upsilon} \; .
\label{bsenergy}
\eqa

\subsection{Finding the excited states}

The basic method for finding excited states is to first evolve the
initially random wavefunction to large imaginary times, find the
ground state wavefunction, $\Psi_0$, and then project this state out
from the initial wavefunction and re-evolve the partial-differential
equation in imaginary time. However, there are (at least) two more
efficient ways to accomplish this. The first is to record snapshots of
the 3d wavefunction at a specified interval $\tau_{\text{snapshot}}$
during a single evolution in $\tau$. After having obtained the ground
state wavefunction, one can go back and extract the excited
states by projecting out the ground state wavefunction from the
recorded snapshots of $\Psi({\bf r},\tau)$ \cite{Sudiarta:2007,Strickland:2009ft}.

An alternative way to select different excited states is to impose a
symmetry condition on the initially random wavefunction which cannot
be broken by the Hamiltonian evolution \cite{Strickland:2009ft}. For example, one can select
the first p-wave excited state by
anti-symmetrizing the initial wavefunction around either the $x$, $y$,
or $z$ axes.  In the non-spherical case this method can be used to
separate the different excited state polarizations in the
quarkonium system and to determine their energy eigenvalues with high
precision.

\section{Application of the sudden approximation}
\label{app:sudden}

In this appendix we explore what happens to a system which suddenly has a magnetic field 
turned on.  We will model this as being instantaneous in order to simplify the treatment
and restrict our attention to a 3d harmonic oscillator eigenstate since it is 
possible to make much more analytic progress in this case.  We start by positing that 
for $t<0$ there is no magnetic field and that the system is subject only to an internal
harmonic interaction in which case the full state can be decomposed in terms of the
no-magnetic-field eigenstates $\Phi^{(0)}_k$
\beq
\Phi(t) = \sum_{k} c_k \Phi^{(0)}_k e^{- i E^{(0)}_k t} \qquad\qquad t < 0 \, ,
\eeq
where $k$ collects all relevant quantum numbers and the sum represents a sum over
discrete quantum numbers and integral for continuous quantum numbers.  
For $t\geq0$ we can expand in terms
of the eigenstates in the presence of the magnetic field $\Phi^{(1)}_m$
\beq
\Phi(t) = \sum_{m} d_m \Phi^{(1)}_m e^{- i E^{(1)}_m t} \qquad\qquad t \geq 0 \, ,
\eeq
At $t=0$ we match the coefficients which requires
\beq
\sum_{m} d_m \Phi^{(1)}_m = \sum_{k} c_k \Phi^{(0)}_k \, .
\eeq
Projecting with $\Phi^{(1)}_n$ and using their orthonormality we obtain
\beq
d_n  = \sum_{k} c_k \langle \Phi^{(1)}_n \!\mid\! \Phi^{(0)}_k \rangle \; ,
\eeq

\subsection{Pure state for $t<0$}

If the state for $t<0$ is a pure state with $c_k = \delta_{km}$ we obtain
\beq
d_n  = \langle \Phi^{(1)}_n \!\mid\! \Phi^{(0)}_m \rangle \, .
\eeq
We now turn to the computation of the overlap integrals necessary for the case at hand.
The $t<0$ states are 
\beq
\Phi^{(0)}_{{\bf P},{n_\perp^0}n_z^0 \ell^0}({\bf R},{\bf r}) = {\cal N}^{(0)} \,   
\rho^{|\ell^0|} e^{i \ell^0 \phi} e^{-\frac{1}{2} \gamma^2 (\rho^2+z^2)} \,
H_{n_z^0}(\gamma z)  L_{n_\perp^0}^{|\ell^0|}(\gamma^2\rho^2) 
e^{i {\bf P} \cdot {\bf R}} \, ,
\eeq
where 
\beq
{\cal N}^{(0)} = \frac{ \gamma^{|\ell^0|+3/2} }{\sqrt{2^{n_z^0} \pi^{3/2}}}
\sqrt{\frac{ n_\perp^0!}{n_z^0! \, (|\ell^0|+n_\perp^0)! }} \; ,
\eeq
and the $t\geq0$ states are
\beq
\Phi^{(1)}_{{\bf K},{n_\perp}n_z \ell}({\bf R},{\bf r}) = {\cal N}^{(1)} \,   
\tilde\rho^{|\ell|} e^{i \ell \tilde\phi} e^{-\frac{1}{2} \gamma^2 z^2}\,e^{-\frac{1}{2}\alpha^2 \tilde\rho^2}\,
H_{n_z}(\gamma z)  L_{n_\perp}^{|\ell|}(\alpha^2\tilde\rho^2) 
e^{i \left( {\bf K} - \frac{1}{2} q {\bf B} \times {\bf r}\right) \cdot {\bf R}} \, ,
\eeq
with
\bqa
\omega_c &=& \frac{qB}{\mu} \, , \nonumber \\
\alpha^2 &=&  \mu \sqrt{ \omega_0^2 + \frac{\omega_c^2}{4} } \, , \nonumber \\
\gamma^2 &=&  \mu \omega_0 \, , \nonumber \\
\tilde\rho^2 &=& (x - \lambda K_y )^2 + (y + \lambda K_x)^2 \, , \nonumber \\
\tilde\phi &=& \arctan\!\left(\frac{y + \lambda K_x}{x - \lambda K_y}\right) , \nonumber \\
\lambda &=& \frac{\omega_c}{4\mu(\omega_0^2 +\omega_c^2/4)} \, ,
\label{eq:definitions}
\eqa
and
\beq
{\cal N}^{(1)} = \frac{ \alpha^{|\ell|+1} \gamma^{1/2} }{\sqrt{2^{n_z} \pi^{3/2}}}
\sqrt{\frac{ n_\perp!}{n_z! \, (|\ell|+n_\perp)! }} \, .
\eeq
The six-dimensional overlap integral in relative cylindrical coordinates becomes
\bqa
d_n  &=& {\cal N}^{(0)} {\cal N}^{(1)} \int_0^\infty \rho \, d \rho \int_0^{2\pi} d\phi \int_{-\infty}^{\infty} dz \int d^3{\bf R} \; \rho^{|\ell^0|} \tilde\rho^{|\ell|} \, e^{i (\ell^0 \phi - \ell \tilde\phi)}
e^{-\gamma^2 z^2}\,e^{-\frac{1}{2}(\gamma^2 \rho^2 + \alpha^2 \tilde\rho^2)}
\nonumber \\
&&  \hspace{3cm}
\times H_{n_z}(\gamma z) H_{n_z^0}(\gamma z) L_{n_\perp^0}^{|\ell^0|}(\gamma^2\rho^2) L_{n_\perp}^{|\ell|}(\alpha^2\tilde\rho^2)
e^{i ({\bf P} - {\bf K} + \frac{1}{2} q {\bf B} \times {\bf r} )\cdot {\bf R}} \, . 
\eqa
Using $\frac{1}{2} q {\bf B} \times {\bf r} = \frac{1}{2} q B (-y,x,0) = \frac{1}{2} q B \rho (-\sin\phi,\cos\phi,0)$ and the orthonormality of the Hermite polynomials we can perform the $z$ and $\bf Z$ integrations.  Using the exponential we can further perform the $\bf X$ and $\bf Y$ integrations.  The remaining two integrals are evaluated in cartesian coordinates.
The result is  
\beq
d_n  =  \tilde {\cal N}_{nm}  \, \left( \frac{2}{|q| B} \right)^2 \, \delta_{n_z n_z^0} \delta(P_z - K_z) 
\rho^{|\ell^0|} \tilde\rho^{|\ell|} \, e^{i (\ell^0 \phi - \ell \tilde\phi)}
\,e^{-\frac{1}{2}(\gamma^2 \rho^2 + \alpha^2 \tilde\rho^2)} 
L_{n_\perp^0}^{|\ell^0|}(\gamma^2\rho^2) L_{n_\perp}^{|\ell|}(\alpha^2\tilde\rho^2) \, ,
\label{eq:dnresult}
\eeq
where
\bqa
\tilde {\cal N}_{nm} &=& (2 \pi)^3 {\cal N}^{(0)} {\cal N}^{(1)} \sqrt\pi \, 2^{n_z} \, n_z!/\gamma 
= 2 (2 \pi)^2 \alpha^{|\ell|+1} \gamma^{|\ell^0|+1} \sqrt{\frac{ n_\perp^0! \, n_\perp! }{(|\ell^0|+n_\perp^0)! \, (|\ell|+n_\perp)!}} \, ,
\nonumber \\
\rho^2 &=& x^2 + y^2 = \left( \frac{2}{qB} \right)^2 \left[ (P_x - K_x)^2 + (P_y - K_y)^2 \right] \, , \nonumber \\
\phi &=& \arctan\!\left(\frac{y}{x}\right) = \arctan\!\left(\frac{P_x - K_x}{K_y - P_y}\right) \, ,
\nonumber \\
\tilde\rho^2 &=& \left( \frac{2}{qB} \right)^2 \left[ \left(\beta K_y - P_y\right)^2 + (P_x - \beta K_x)^2 \right] \, , \nonumber \\
\tilde\phi &=& \arctan\!\left(\frac{P_x - \beta K_x}{\beta K_y - P_y}\right) \, ,
\eqa
with
\beq
\beta \equiv \frac{8 \omega_0^2 + \omega_c^2}{8 \omega_0^2 + 2 \omega_c^2} \; ,
\eeq
which satisfies $\frac{1}{2} \leq \beta \leq 1$.  Note that the above definitions only apply for the probability amplitude $d_n$.  For $\tilde\rho$ and $\tilde\phi$ in the wavefunction, we need to use the definitions in Eq.~(\ref{eq:definitions}).

\subsection{Gaussian Wave Packet as Initial Condition}

Let's consider that the initial condition is not a pure state but instead a Gaussian linear combination
\beq
\Phi(t) = \sum_k c_k \Phi_k^{(0)} e^{-i E_k^{(0)} t} \, ,
\eeq
where $k = (\ell, k_z, k_\perp,{\bf P})$.
We will assume that the system is in a well-defined internal state $(\ell^0,n_z^0,n_\perp^0)$ but has a spread in COM momentum:
\beq
c_k = \sqrt{\frac{8 \pi^{3/2}}{\sigma^3}} \delta_{\ell^0 \ell} \delta_{n_z^0 k_z} \delta_{n_\perp^0 k_\perp} e^{-({\bf P} - {\bf P}^0)^2/(2 \sigma^2)} \, .
\eeq
In this case the coefficient $d_n$ is more complicated:
\beq
d_n  = \sum_{m} c_m \langle \Phi^{(1)}_n \!\mid\! \Phi^{(0)}_m \rangle \, ,
\eeq
where we can use the pure state result obtained previously
\beq
\langle \Phi^{(1)}_n \!\mid\! \Phi^{(0)}_m \rangle  =  \tilde {\cal N}_{nm}  \, \left( \frac{2}{|q| B} \right)^2 \, \delta_{n_z n_z^0} \delta(P_z - K_z) 
\rho^{|\ell^0|} \tilde\rho^{|\ell|} \, e^{i (\ell^0 \phi - \ell \tilde\phi)}
\,e^{-\frac{1}{2}(\gamma^2 \rho^2 + \alpha^2 \tilde\rho^2)} 
L_{n_\perp^0}^{|\ell^0|}(\gamma^2\rho^2) L_{n_\perp}^{|\ell|}(\alpha^2\tilde\rho^2) \, ,
\label{eq:dnresultg}
\eeq
with $m = (\ell^0,n_z^0,n_\perp^0,{\bf P})$ and $n = (\ell,n_z,n_\perp,{\bf K})$.

\subsection{Time evolution of the center-of-mass kinetic momentum}

We consider next the evolution of the COM kinetic momentum after the magnetic field is applied. We seek to evaluate $\langle {\bf P}_{\rm kinetic} \rangle = \langle\Phi(t)|{\bf P}_{\rm kinetic}|\Phi(t)\rangle$ for $t>0$.
\beq
\langle\Phi(t)|{\bf P}_{\rm kinetic}|\Phi(t)\rangle = \sum_{m,n} d_m^* d_n \langle\Phi_m^{(1)}|{\bf P}_{\rm kinetic}|\Phi_n^{(1)}\rangle e^{- i (E^{(1)}_n - E^{(1)}_m) t} \, ,
\eeq
where $m = ({\ell^\prime ,n_z^\prime,n_\perp^\prime,{\bf K}^\prime})$, $n = ({\ell,n_z,n_\perp},{\bf K})$, and
\bqa
\sum_{m} &\equiv& 
\sum_{n_z^\prime=0}^\infty \sum_{\ell^\prime =-\infty}^\infty \sum_{n_\perp^\prime=0}^\infty 
\int\!\frac{d^3{\bf K}^\prime}{(2\pi)^3} \, ,
\nonumber \\
\sum_{n} &\equiv& 
\sum_{n_z=0}^\infty \sum_{\ell=-\infty}^\infty \sum_{n_\perp=0}^\infty 
\int\!\frac{d^3{\bf K}}{(2\pi)^3} \, .
\eqa
\bqa
\langle\Phi_m^{(1)}|{\bf P}_{\rm kinetic}|\Phi_n^{(1)}\rangle 
&=& \langle\Phi_m^{(1)}|{\bf K} - q {\bf B} \times {\bf r} |\Phi_n^{(1)}\rangle
\nonumber \\
&=& {\bf K} \delta_{mn} - q B \langle\Phi_m^{(1)}| \left(-\tilde\rho\sin\tilde\phi + \frac{c}{a},\tilde\rho\cos\tilde\phi + \frac{b}{a} ,0\right) |\Phi_n^{(1)}\rangle \, ,
\eqa
where $\delta_{mn} = \delta_{\ell^\prime \ell} \, \delta_{n_z^\prime n_z} \, \delta_{n_\perp^\prime n_\perp} \, \delta_{\bf K^\prime {\bf K}}$, $\delta_{\bf K^\prime {\bf K}} \equiv (2\pi)^3 \delta^3({\bf K}^\prime-{\bf K})$, and we remind the reader that $a = \mu (\omega_0^2 +\omega_c^2/4)$, $b =  \omega_c K_y/4$, $c =  \omega_c K_x/4$.
Considering the second term we have
\beq
 \left(-\langle\Phi_m^{(1)}|\,\tilde\rho\sin\tilde\phi\,|\Phi_n^{(1)}\rangle + \frac{c}{a} \delta_{mn},\langle\Phi_m^{(1)}|\,\tilde\rho\cos\tilde\phi\,|\Phi_n^{(1)}\rangle + \frac{b}{a} \delta_{mn} ,0\right) .
\eeq
To proceed, we first consider 
\beq
J_{mn}^+ \equiv \langle\Phi_m^{(1)}|\tilde\rho \, e^{i\tilde\phi} \, |\Phi_n^{(1)}\rangle \; ,
\eeq
and
\beq
J_{mn}^- \equiv \langle\Phi_m^{(1)}|\tilde\rho \, e^{-i\tilde\phi} \, |\Phi_n^{(1)}\rangle \, .
\eeq
For $\ell \geq 0$
\beq
J_{mn}^+ = \frac{\delta_{\bf K^\prime {\bf K}}\delta_{n_z^\prime n_z} \delta_{\ell^\prime ,\ell+1}}{\alpha}
\left[  \delta_{n_\perp^\prime n_\perp} \sqrt{n_\perp+\ell+1}
- \delta_{n_\perp^\prime,n_\perp-1} \sqrt{n_\perp}
\right] \, .
\eeq
For $\ell \leq -1$
\beq
J_{mn}^+ = \frac{\delta_{\bf K^\prime {\bf K}}\delta_{n_z^\prime n_z} \delta_{\ell^\prime ,\ell+1}}{\alpha}
\left[  \delta_{n_\perp^\prime n_\perp} \sqrt{n_\perp-\ell}
- \delta_{n_\perp^\prime,n_\perp+1} \sqrt{n_\perp+1}
\right] \, .
\eeq
For $\ell \geq 1$
\beq
J_{mn}^- = \frac{\delta_{\bf K^\prime {\bf K}}\delta_{n_z^\prime n_z} \delta_{\ell^\prime ,\ell-1}}{\alpha}
\left[  \delta_{n_\perp^\prime n_\perp} \sqrt{n_\perp+\ell}
- \delta_{n_\perp^\prime,n_\perp+1} \sqrt{n_\perp+1}
\right] \, .
\eeq
For $\ell \leq 0$
\beq
J_{mn}^- = \frac{\delta_{\bf K^\prime {\bf K}}\delta_{n_z^\prime n_z} \delta_{\ell^\prime ,\ell-1}}{\alpha}
\left[  \delta_{n_\perp^\prime n_\perp} \sqrt{n_\perp-\ell+1}
- \delta_{n_\perp^\prime,n_\perp-1} \sqrt{n_\perp}
\right] \, .
\eeq
With these we have determined
\bqa
\langle\Phi_m^{(1)}|\, \tilde\rho\sin\tilde\phi \,|\Phi_n^{(1)}\rangle 
&=& \frac{1}{2i} (J_{mn}^+ - J_{mn}^-) \equiv {\cal S}_{mn} \, ,
\nonumber \\
\langle\Phi_m^{(1)}|\, \tilde\rho\cos\tilde\phi \,|\Phi_n^{(1)}\rangle 
&=& \frac{1}{2} (J_{mn}^+ + J_{mn}^-) \equiv {\cal C}_{mn} \, .
\eqa
To evaluate $\langle\Phi(t)|{\bf P}_{\rm kinetic}|\Phi(t)\rangle$ we will need
\bqa
&&
\sum_{\ell=-\infty}^\infty 
\sum_{n_\perp=0}^\infty
\sum_{\ell^\prime =-\infty}^\infty 
\sum_{n_\perp^\prime=0}^\infty 
\int_{{\bf K}^\prime}
d_m^* d_n J^{\pm}_{mn} \, e^{- i (E^{(1)}_n - E^{(1)}_m) t}
\nonumber \\
&&
=
\frac{1}{\alpha}
\sum_{\ell=0}^{\infty} 
\sum_{n_\perp=0}^\infty
\left[ d_{\pm(\ell+1),n_\perp}^* d_{\pm\ell,n_\perp} \sqrt{n_\perp+\ell+1} \; e^{i \alpha^2 t/\mu}
- d_{\pm(\ell+1),n_\perp-1}^* d_{\pm\ell,n_\perp} \sqrt{n_\perp} \; e^{-i \alpha^2 t/\mu}
\right]
\nonumber \\
&&
+
\frac{1}{\alpha}
\sum_{\ell=1}^{\infty} 
\sum_{n_\perp=0}^\infty
\left[  d_{\mp(\ell-1),n_\perp}^* d_{\mp\ell, n_\perp}\sqrt{n_\perp+\ell} \; e^{-i \alpha^2 t/\mu}
- d_{\mp(\ell-1),n_\perp+1}^* d_{\mp\ell, n_\perp}\sqrt{n_\perp+1} \; e^{i \alpha^2 t/\mu}
\right] \, ,
\nonumber \\
\eqa
where we have used Eq.~(\ref{eq:harmen}).

To proceed we note that
$d_{\ell,n_\perp}^* d_{\ell^\prime ,n_\perp^\prime} = d_{-\ell^\prime ,n_\perp^\prime}^*d_{-\ell,n_\perp}$.  
Now we have after some work
\bqa
&&
\sum_{\ell=-\infty}^\infty 
\sum_{n_\perp=0}^\infty
\sum_{\ell^\prime =-\infty}^\infty 
\sum_{n_\perp^\prime=0}^\infty 
\int_{{\bf K}^\prime}
d_m^* d_n (J^+_{mn}\pm J^-_{mn}) \, e^{- i (E^{(1)}_n - E^{(1)}_m) t}
\nonumber \\
&& 
\hspace{2cm}
= 
\frac{2}{\alpha}
\sum_{\ell=0}^{\infty} 
\sum_{n_\perp=0}^\infty
\big[ ( d_{-\ell,n_\perp}^* d_{-\ell -1,n_\perp}\pm d_{\ell,n_\perp}^*d_{\ell +1,n_\perp}) \sqrt{n_\perp+\ell+1}
\nonumber \\
&&
\hspace{3cm}
- (d_{-\ell,n_\perp+1}^* d_{-\ell -1,n_\perp}\pm d_{\ell,n_\perp+1}^* d_{\ell +1,n_\perp}) \sqrt{n_\perp+1} \;
\big]\cos(\alpha^2 t/\mu) \; . 
\eqa
Using $d_{\ell,n_\perp} = \int d^2 {\bf P}_\perp \, m_{\ell,n_\perp}$ with
\bqa
m_{\ell,n_\perp} &=& \frac{\tilde {\cal N}}{(2\pi)^3}  \, \sqrt{\frac{8 \pi^{3/2}}{\sigma^3}} \, \left( \frac{2}{|q| B} \right)^2 \delta_{n_z^0 n_z} e^{-(K_z - P^0_z)^2/(2\sigma^2)}
\nonumber \\
&& \hspace{1mm} \times 
e^{-({\bf P}_\perp - {\bf P}^0_\perp)^2/(2 \sigma^2)}
\rho^{|\ell^0|} \tilde\rho^{|\ell|} \, e^{i (\ell^0 \phi - \ell \tilde\phi)}
\,e^{-\frac{1}{2}(\gamma^2 \rho^2 + \alpha^2 \tilde\rho^2)} 
L_{n_\perp^0}^{|\ell^0|}(\gamma^2\rho^2) L_{n_\perp}^{|\ell|}(\alpha^2\tilde\rho^2) \; ,
\eqa
\bqa
\tilde {\cal N} &=& 2 (2 \pi)^2 \alpha^{|\ell|+1} \gamma^{|\ell^0|+1} \sqrt{\frac{ n_\perp^0! \, n_\perp! }{(|\ell^0|+n_\perp^0)! \, (|\ell|+n_\perp)!}} \; ,
\eqa
and a recurrence relation for the Laguerre polynomials we can write
\bqa
&&
\sum_{\ell=-\infty}^\infty 
\sum_{n_\perp=0}^\infty
\sum_{\ell^\prime =-\infty}^\infty 
\sum_{n_\perp^\prime=0}^\infty 
\int_{{\bf K}^\prime}
d_m^* d_n (J^+_{mn}\pm J^-_{mn}) \, e^{- i (E^{(1)}_n - E^{(1)}_m) t}
\nonumber \\
&& 
\hspace{2cm}
= 
2\cos(\alpha^2 t/\mu)
\sum_{\ell=0}^{\infty} 
\sum_{n_\perp=0}^\infty
\big( d_{-\ell,n_\perp}^*\int d^2{\bf P_\perp} \, m_{-\ell -1,n_\perp}\tilde\rho \, e^{i\tilde\phi}
\pm d_{\ell,n_\perp}^*\int d^2{\bf P_\perp} \, m_{\ell,n_\perp}\tilde\rho \, e^{-i\tilde\phi}\big)
\nonumber \\
&& 
\hspace{2cm}
= 
2\cos(\alpha^2 t/\mu)
\sum_{\ell=0}^{\infty} 
\sum_{n_\perp=0}^\infty
d_{\ell,n_\perp}^*\int d^2{\bf P_\perp} \, m_{\ell,n_\perp}\tilde\rho 
\big( e^{i\tilde\phi}\pm e^{-i\tilde\phi}\big) \; .
\eqa
With this we can obtain
\bqa
&&
\sum_{\ell=-\infty}^\infty 
\sum_{n_\perp=0}^\infty
\sum_{\ell^\prime =-\infty}^\infty 
\sum_{n_\perp^\prime=0}^\infty 
\int_{{\bf K}^\prime}
d_m^* d_n {\cal C}_{mn} \, e^{- i (E^{(1)}_n - E^{(1)}_m) t}
\nonumber \\
&&
\hspace{4cm}
= 
2\cos(\alpha^2 t/\mu)
\sum_{\ell=0}^{\infty} 
\sum_{n_\perp=0}^\infty
d_{\ell,n_\perp}^*\int d^2{\bf P_\perp} \, m_{\ell,n_\perp}\tilde\rho \, \cos\tilde\phi \; ,
\eqa
and
\bqa
&&
\sum_{\ell=-\infty}^\infty 
\sum_{n_\perp=0}^\infty
\sum_{\ell^\prime =-\infty}^\infty 
\sum_{n_\perp^\prime=0}^\infty 
\int_{{\bf K}^\prime}
d_m^* d_n {\cal S}_{mn} \, e^{- i (E^{(1)}_n - E^{(1)}_m) t}
\nonumber \\
&&
\hspace{4cm}
=
2\cos(\alpha^2 t/\mu)
\sum_{\ell=0}^{\infty} 
\sum_{n_\perp=0}^\infty
d_{\ell,n_\perp}^*\int  d^2{\bf P_\perp} \, m_{\ell,n_\perp}\tilde\rho \, \sin\tilde\phi \; .
\eqa

Recall we are after
\bqa
\langle{\bf P}_{\rm kinetic}\rangle &=& 
\sum_{n_z=0}^\infty  
\sum_{n_\perp=0}^\infty  
\sum_{\ell=-\infty}^\infty 
\int\!\frac{d^3{\bf K}}{(2\pi)^3} \, d_n^* d_n \left[{\bf K} - q B\left(\frac{c}{a},\frac{b}{a},0\right)\right]
\nonumber \\
&& \hspace{-8mm} 
- \; q B
\sum_{n_z^\prime=0}^\infty
\sum_{n_\perp^\prime=0}^\infty
\sum_{\ell^\prime =-\infty}^\infty
\int\!\frac{d^3{\bf K^{\prime}}}{(2\pi)^3} \,
d_m^* d_n 
\left(-{\cal S}_{mn},{\cal C}_{mn},0\right) e^{- i (E^{(1)}_n - E^{(1)}_m) t} \, .
\eqa
Using what we just learned we have
\bqa
\langle{\bf P}_{\rm kinetic}\rangle &=& 
\sum_{n_z=0}^\infty  
\sum_{n_\perp=0}^\infty  
\sum_{\ell=-\infty}^\infty 
\int\!\frac{d^3{\bf K}}{(2\pi)^3} \, d_n^* d_n \left[{\bf K} - q B\left(\frac{c}{a},\frac{b}{a},0\right)\right]
\nonumber \\
&& \hspace{-8mm} 
- \; 2 q B \cos(\alpha^2 t/\mu)
\sum_{n_z=0}^\infty
\sum_{n_\perp=0}^\infty  
\sum_{\ell=0}^\infty
\int\!\frac{d^3{\bf K}}{(2\pi)^3} \,
d_{\ell,n_\perp}^*\int d^2{\bf P_\perp} \, m_{\ell,n_\perp}\tilde\rho \,
\big(-\sin\tilde\phi \, , \cos\tilde\phi \, , 0\big) \; .
\nonumber \\
\eqa
Focusing on the second term, we need to evaluate
\bqa
\sum_{n_z=0}^\infty
\sum_{n_\perp=0}^\infty  
\sum_{\ell=0}^\infty
\int\!\frac{d^3{\bf K}}{(2\pi)^3} \;
d_{\ell,n_\perp}^*\int d^2{\bf P_\perp} \, m_{\ell,n_\perp}\tilde\rho \,
\big(-\sin\tilde\phi \, , \cos\tilde\phi \, , 0\big) \;.
\eqa

The summation over $n_z$ and integration over $K_z$ can be done analytically.  Next, we change integration variables from $({\bf K_\perp } , {\bf P_\perp } )$ to $(\rho ,\phi ,\tilde\rho ,\tilde\phi )$ and use the completeness of the Lagueere polynomials to eliminate the summation over $n_\perp$.  Now, one of the integrals over $\tilde\rho $ and the summation over $\ell$ can be done analytically.  The remaining five integrals are evaluated numerically and found to converge to zero.  We now have
\bqa
&&
\langle{\bf P}_{\rm kinetic}\rangle =
\sum_{n_z=0}^\infty  
\sum_{n_\perp=0}^\infty  
\sum_{\ell=-\infty}^\infty 
\int\!\frac{d^3{\bf K}}{(2\pi)^3} \, d_n^* d_n \left[{\bf K} - q B\left(\frac{c}{a},\frac{b}{a},0\right)\right] 
\nonumber \\  && \hspace{1.75cm}
= \frac{4\omega_0^2}{4\omega_0^2+\omega_c^2}
\sum_{n_z=0}^\infty  
\sum_{n_\perp=0}^\infty  
\sum_{\ell=-\infty}^\infty 
\int\!\frac{d^3{\bf K}}{(2\pi)^3} \, d_n^* d_n {\bf K_\perp}
\nonumber \\  && \hspace{3cm}
+ \; \hat z
\sum_{n_z=0}^\infty
\sum_{n_\perp=0}^\infty  
\sum_{\ell=-\infty}^\infty 
\int\!\frac{d^3{\bf K}}{(2\pi)^3} \, d_n^* d_n K_z \; .
\eqa

Again, the summation over $n_z$ and integration over $K_z$ can be done analytically.  We change variables, use the completeness of the Lagueere polynomials, and do one of the integrals over $\tilde\rho$.  Now, we use the completeness of the azimuthal modes and do one of the integrals over $\tilde\phi$.
\bqa
&&
\langle{\bf P}_{\rm kinetic}\rangle = \left(\frac{\lambda}{\pi \sigma}\right)^2 \gamma^{2(|\ell^0|+1)}\frac{4\omega_0^2}{4\omega_0^2+\omega_c^2} \frac{n_\perp^0!}{(|\ell^0|+n_\perp^0)!}
 \int \! \rho d\rho \int \! d\phi
\nonumber \\ 
&& \hspace{20mm} \times
\rho^{2|\ell^0|} \, e^{-\gamma^2 \rho^2}
 \left(L_{n_\perp^0}^{|\ell^0|}(\gamma^2\rho^2)\right)^2
\int \! \tilde\rho d\tilde\rho \int \! d\tilde\phi \;
{\bf K}_\perp e^{-({\bf P}_\perp - {\bf P}^0_\perp)^2/\sigma^2}
\nonumber \\
&& \hspace{15mm}
+ \; \hat z \; \left(\frac{\lambda}{\pi \sigma}\right)^2 \gamma^{2(|\ell^0|+1)}P_z^0 \frac{n_\perp^0!}{(|\ell^0|+n_\perp^0)!}
 \int \! \rho d\rho \int \! d\phi
\nonumber \\ 
&& \hspace{20mm} \times
\rho^{2|\ell^0|} \, e^{-\gamma^2 \rho^2}
 \left(L_{n_\perp^0}^{|\ell^0|}(\gamma^2\rho^2)\right)^2
\int \! \tilde\rho d\tilde\rho \int \! d\tilde\phi \;
e^{-({\bf P}_\perp - {\bf P}^0_\perp)^2/\sigma^2} \; .
\eqa
Using
\bqa
&&
P_x = -\lambda (\beta y - \tilde y) \; ,
\nonumber \\ &&
P_y = \lambda (\beta x - \tilde x) \; ,
\nonumber \\ &&
K_x=-\lambda(y-\tilde y) \; ,
\nonumber \\ &&
K_y=\lambda(x-\tilde x) \; ,
\eqa
and the orthogonality of the Laguerre polynomials, the remaining integrals can be done analytically.
The final result is
\bqa
\langle{\bf P}_{\rm kinetic}\rangle &=& \left(\frac{4\omega_0^2}{4\omega_0^2+\omega_c^2}\;P_x^0\;,\;\frac{4\omega_0^2}{4\omega_0^2+\omega_c^2}\;P_y^0\;,\;P_z^0\right)\;.
\eqa
%

\bibliography{quarkonium}

\end{document}